\documentclass[a4paper,showpacs,amsmath,amssymb,12pt]{article}
\usepackage{amscd}
\usepackage{mathrsfs}
\usepackage{amsmath}
\usepackage{amsfonts}
\usepackage{indentfirst}
\usepackage{amssymb}
\usepackage{latexsym}
\usepackage{graphicx}
\usepackage[english]{babel}
\usepackage{epsfig}
\usepackage{graphicx}
\usepackage{slashbox}
\input epsf.sty

\title{Production of Triply Heavy Baryons at LHC}
\author{Yu-Qi Chen and Su-Zhi Wu \\
{\small Key Laboratory of Frontiers in Theoretical
Physics,}\\
{\small Institute of Theoretical Physics,Chinese Academy of
Sciences} \\
{\small Beijing 100190, P.R. China}}
\date{}
\begin{document}
\maketitle
\begin{abstract}
Triply heavy baryons are very interesting hadrons to be explored for
they provide particular information about strong interactions,
hadron structures, and weak decays of heavy baryons. We calculate
the hadronic production cross sections of  the $\Omega_{ccc}$ and
the $\Omega_{ccb}$, which are dominated by the $gg$ fusion
subprocesses containing 4362 and 1454 Feynman diagrams,
respectively. A method for generating and calculating tree level
Feynman diagrams automatically is developed to deal with complicated
processes containing so many diagrams. Our results show that
$10^4-10^5$ events of triply heavy baryons can be accumulated for 10
$fb^{-1}$ integrated luminosity at LHC. Signatures of the triply
heavy baryons are pointed out, with emphasis on the decay modes
$\Omega_{ccc} \to \Omega_{sss} + 3 \pi^+$ and $\Omega_{ccb} \to
\Omega_{sss} + 3 \pi^+ +\pi^-$. We conclude that it is quite
promising to discover them at LHC.
\end{abstract}

\newpage
\section{Introduction}

One interesting baryon to be discovered is the triply heavy baryon
which is the color-singlet bound state consisting of three heavy $c$
or $b$ quarks by the strong interaction. Especially, exploring this
baryon will be very useful for understanding the three body static
potential. Since the mass $m$ of the heavy quark is so high, the
relative motion of the quarks inside the triply heavy baryon is
nonrelativistic and the typical velocity $v$ of the heavy quark in
the rest frame of the baryon is small. The three heavy quarks are
bound tightly and the typical size of this baryon being order of
$1/mv$ is smaller than that of the conventional baryons with light
flavors. The mass of the baryon $\overline{M}$ is the sum of the
three heavy quarks' masses and the binding energy which is of order
$mv^2$. Similar with the heavy quarkonium system, in the limit of $v
\ll 1$, there are three distinct energy scales involved in the
triply heavy baryon system, i.e., the mass of heavy quark $m$, the
three-momentum of the heavy quark $mv$, and the off-shell energy of
the heavy quark $mv^2$. Nonrelativistic QCD effective theory (NRQCD)
\cite{xNRQCD} can be used to describe the triply heavy baryon system
in a simpler way by reducing the number of energy scales, where
physics effects at energy scale $m$ are explicitly integrated out.
Moreover, in the limits of both large $N_c$ and heavy quark mass,
the mean field approximation can be applied to this system.
Consequently, the wave function satisfies an ordinary differential
equation \cite{witten}. Exploring this baryon certainly will enrich
our knowledge about hadron structures and QCD interactions. In
addition, the ground state can only decay via the weak interaction.
Since the dynamics is relatively simpler, the triply heavy baryon
provides very interesting samples to study the weak decays of
hadrons. In particular, both the identical fermion effects and the
larger recoil effects caused by the spectators can be studied well
in theory. It may provide some insight for clarifying some
long-standing puzzles in weak decays of heavy hadrons.

The masses, lifetimes, and other properties of these baryons have
been studied over the past three decades
\cite{xm1,xm2,xm3,xm4,xm5,Bfir,x1,x2}. However, productions of these
baryons are so difficult that they have not yet been discovered so
far. They can be produced at high energy colliders by a direct
production mechanism which can be described as follows. Triple heavy
quark pairs are produced first at energy scale $m$ or higher,
followed by the formation of the triply heavy baryons at energy
scale $mv$. In the leading order approximation, three heavy quarks
forming a color singlet state move together with smaller relative
velocities $v$. Therefore, the production cross sections can be
factored into short-distance coefficients and long-distance matrix
elements. The short-distance coefficients describe the hard process
effects of the triple heavy quark pairs production which can be
calculated by perturbative QCD and expanded as a power series of
$\alpha_s$ at energy scale $m$ or higher. The long-distance matrix
elements describe the formation of the triply heavy baryon from
point-like three heavy quarks with small relative velocity $v$. At
hadron colliders, the triple heavy quark pairs can be produced via
$gg$ fusion and quark-antiquark annihilation subprocess while at
$e^+e^-$ colliders, it can be produced via $e^+e^-$ annihilation
process. The production cross section for the triply-charmed baryon
in the $e^+e^-$ collision has been calculated in Ref. \cite{x5} and
the predicted production rate turns out to be very small. LHC,
however, as a very high energy and high luminosity hadron collider,
provides a very good chance to discover this sort of baryons due to
very large heavy quark pairs production rates.

The authors in Refs. \cite{frag-1} and \cite{frag-2} calculated the
hadronic production cross sections of the triply heavy baryons at
LHC with their fragmentation functions. They derived fragmentation
functions in the so-called diquark model of the heavy baryon and in
perturbative calculation model, respectively. However, their
obtained results do not represent the fragmentation functions for
the production of triply heavy baryons predicted by QCD perturbation
theory in the leading order.

In \cite{frag-1}, the author calculated the fragmentation function
in the diquark model of the triply heavy baryons by treating two of
the three heavy quarks as a point-like diquark. This is
inappropriate for the reason that two heavy quarks can be treated as
a point-like diquark particle only when the momentum of the gluon
emitted or absorbed from the diquark is much smaller than the
inverse power of the typical size of the diquark which is order of
$mv$. In the production process, this condition cannot be satisfied
since the momenta of those gluons are order of $m$ or higher.
Moreover, in his diquark description, the produced two free heavy
antiquarks are forced to move in the same direction, which deviates
from QCD description.

  In Ref. \cite{frag-2}, the authors calculated the fragmentation function
using perturbation theory in  Feynman gauge and  including
contributions only from two Feynman diagrams.  As shown in Ref.
\cite{Chen:1993ii}, the fragmentation function can only be correctly
calculated in axial gauge. Moreover, there are seven Feynman
diagrams contributing to the leading order fragmentation function
which are shown in Fig.1 in this paper.

Even with the correct QCD leading order fragmentation function, one
cannot expect that it gives accurate description for the triply
heavy baryons at small $P_T$. In Ref. \cite{chang-chen-h}, the
authors did a comparative studied for the hadronic $B_c$ production
by full QCD and by the fragmentation approximation. Their results
show that the fragmentation approximation is valid only when the
transverse momentum of the heavy hadron, $P_T$, is much larger than
its mass $\overline{M}$. However, in hadron collisions, the
production cross sections are dominated in the smaller $P_T$ region,
typically comparable  with $\overline{M}$. Thus one can not expect
that the fragmentation approximation gives accurate results.
Instead, calculations of the production cross sections by full
perturbative QCD are  necessary and expected to give reliable
predictions.

In this paper, we study the production rates of the triply heavy
baryons at LHC. We first calculate the short-distance coefficients
in the leading order of $\alpha_s^6$ corresponding to the tree level
contributions. Even at the tree level, there are a number of Feynman
diagrams that need to be calculated and it turns out to be hard
work. For instance, there are as many as 4362 Feynman diagrams for
$gg \to \Omega_{ccc} + \bar{c}\;\bar{c}\;\bar{c} $ and as many as
1454 Feynman diagrams for $gg \to \Omega_{ccb} +
\bar{c}\;\bar{c}\;\bar{b} $.  It is impossible to carry out the
calculations by conventional calculation methods for cross sections.

For a simpler production process $e^+e^- \to \Omega_{ccc} +
\bar{c}\;\bar{c}\;\bar{c} $, to simplify the calculation, the
authors in Ref. \cite{x5} took an approximation ignoring the heavy
quark mass terms in the numerator of the quark propagator. The
approximation significantly simplifies the calculations. However, as
pointed out in Ref. \cite{Chang:1992bb}, the approximation will lead
to quite large errors  with the same order contributions.

To overcome the difficulty meeting here, we propose a method to
evaluate Feynman diagrams by  generating the diagrams automatically
and use a direct amplitude calculation of the Feynman diagrams. The
amplitudes are classified as a set of gauge invariant subsets in
terms of independent color structures. Consequently, the code for
numerical calculations is written in a significant compact form and
numerical calculations of the amplitudes can be carried out. To
ensure the correctness of the calculations, we verify various gauge
invariance of the amplitude of each subset  to examine the code.

Taking the long-distance matrix elements given in \cite{xm4}, the
cross sections of the subprocesses can be calculated. Incorporating
with the parton model, we calculate the production cross sections of
the triply heavy baryons at LHC. Our results show that $10^4-10^5$
events of triply heavy baryons $\Omega_{ccc}$ and $\Omega_{ccb}$ can
be accumulated for 10 $fb^{-1}$ integrated luminosity at LHC. We
also present $P_T$-distributions and rapidity distributions with
various parameters. In this paper, we consider only the ground
states with all orbital angular momenta vanishing. We also point out
the signatures to reconstruct the triply heavy baryons, with
$\Omega_{ccc} \to \Omega_{sss} + 3 \pi^+$ and $\Omega_{ccb} \to
\Omega_{sss} + 3 \pi^+ +\pi^-$ being emphasized. We conclude that it
is quite promising to discover them at LHC.

The rest of the paper is organized as follows. In Sec 2, we present
the calculation of the subprocesses of $gg\;(q\bar{q}) \to
\Omega_{Q_1Q_2Q_3}\bar{Q}_1\bar{Q}_2\bar{Q}_3 $. In Sec. 3,
numerical results of the total cross sections and differential
distributions are presented. Sec. 4 contributes to the signatures of
these baryons and conclusions.

\section{Subprocess of
 $gg\; (q\bar{q})\to \Omega_{Q_1Q_2Q_3}\bar{Q}_1\bar{Q}_2\bar{Q}_3$}

In this section, we calculate the cross sections of the production
subprocesses of $\Omega_{Q_1Q_2Q_3}$  at LHC in the leading order.
There are a number of Feynman diagrams contributing to the short
distance coefficients. We carry out the calculations by taking a
strategy to generating the diagrams automatically and using a direct
amplitude calculations of the Feynman diagrams. Consequently, the
code for numerical calculations is written in a very compact form.
We present the calculations in detail below.

\subsection{Production mechanism of the triply heavy baryon}

The triply heavy baryon is the color-singlet bound state of three
heavy quarks by the strong interaction.  However, since the mass $m$
of the heavy quark is much larger than $\Lambda_{QCD}$, the relative
motion of the quarks inside the triply heavy baryon is
nonrelativistic and the typical velocity $v$ of the heavy quark in
the rest frame of the baryon is small. The three heavy quarks are
bound tightly and the typical size of this baryon being of order
$1/mv$ is smaller than that of the conventional baryons with light
flavors which is of order $1/\Lambda_{QCD}$. The mass of the state
is the sum of the three heavy quark masses and the binding energy
which is  of order $mv^2$. In the limit of $v \ll 1$, there are
three distinct energy scales involved in the triply heavy system,
i.e., the mass of the heavy quark $m$, the three-momentum of the
heavy quark $mv$, and the off-shell energy of the heavy quark
$mv^2$. NRQCD \cite{xNRQCD}  can be used to describe the system
conveniently, where physics effects at energy scale $m$ are
explicitly integrated out.

The triply heavy baryon, $\Omega_{Q_1Q_2Q_3}$, being in a  color
singlet state implies that its color wavefunction must be of the
form $\frac{1}{\sqrt{6}}
\varepsilon^{\xi_1\xi_2\xi_3}Q_{1\xi_1}Q_{2\xi_2}Q_{3\xi_3}$, where
$\xi_i$ ($i$=1,2,3) are the color indices of the heavy quark $Q_i$.
The heavy quarks may be either the $c$ quark or the $b$ quark. The
triply heavy baryons can be classified as two classes in terms of
containing two or three  identical heavy quarks.  For those orbital
angular momentum ground states, the exchange antisymmetry of the
identical fermion implies that the triply heavy baryons with three
identical heavy quarks can only be the spin-symmetrical states and
only spin $\frac{3}{2}$ states are allowed, while the triply heavy
baryons with double identical heavy quarks can be either the
spin-symmetrical state with spin $3/2$ or the other one with spin
$1/2$. We take $\Omega_{ccc}$, $\Omega^*_{ccb}$ and $\Omega_{ccb}$
to denote the triply heavy baryons consisting of the $ccc$ quarks
with spin $3/2$, the $ccb$ quarks with spin $3/2$ and spin $1/2$,
respectively.

The mass of the triply heavy baryon is so high that it is very
difficult to produce at normal high energy machines. At LHC, they
can be produced by a direct production mechanism in which triple
heavy quark pairs are produced first at energy scale $m$ or higher
by $gg$ fusion or $q\bar{q}$ annihilation subprocesses, followed by
the formation of the triply heavy baryons by combining the
color-singlet triple heavy quarks moving in the same directions with
small relative velocities at energy scale $mv$. In the heavy quark
limit, $v \ll 1$, the production cross sections can be factored into
the product of short-distance coefficients and the long-distance
matrix elements. The short-distance coefficients describe the hard
process effects of the triple heavy quark pairs production. They can
be calculated by perturbative QCD and be expanded as a power series
of $\alpha_s$ at energy scale $m$ or higher. The long-distance
matrix elements describe the formation of the triply heavy baryons
from point-like three heavy quarks with small relative velocities.
In this paper, we consider only the contributions arising from the
leading order short distance coefficient and the leading order
long-distance matrix element. There is only single long-distance
matrix element. The factorization then holds at amplitude level
which we illustrate below. Given the leading Fock state description
presented in Appendix \ref{non-baryon}, by the standard perturbation
theory,  the amplitude of the subprocess $gg\ \ (q\bar{q})\to
\Omega_{Q_1Q_2Q_3}\bar{Q}_1\bar{Q}_2\bar{Q}_3$ reads:


\begin{eqnarray}\label{total}
A(gg \ \
(q\bar{q})\to\Omega_{Q_1Q_2Q_3}\bar{Q}_1\bar{Q}_2\bar{Q}_3)&=& \int
\frac{d^3V_1}{(2\pi)^3}\frac{d^3V_2}{(2\pi)^3}
\frac{\sqrt{2\overline{M}}}{\sqrt{8m_1\,m_2\,m_3}}
\nonumber \\
\frac{\psi(\overrightarrow{V}_1,\overrightarrow{V}_2)} {\sqrt{d!}}
&\times& M(gg \ \ (q\bar{q})\to
(Q_1Q_2Q_3)_1^{(s,s_Z)}\bar{Q}_1\bar{Q}_2\bar{Q}_3)\;,
\end{eqnarray}
where $\overline{M}$ ($\overline{M}$=$m_1$+$m_2$+$m_3$) is the mass
of the baryon $\Omega_{Q_1Q_2Q_3}$, $m_i$ is the mass of $Q_i$
(i=1,2,3), $\psi(\overrightarrow{V}_1,\overrightarrow{V}_2)$ is the
wave function of the baryon $\Omega_{Q_1Q_2Q_3}$ in momentum space;
$\overrightarrow{V}_1$, $\overrightarrow{V}_2$ and
-$(\overrightarrow{V}_1+\overrightarrow{V}_2)$ are the relative
momenta among the three quarks in this baryon; $d=2,3$ is the number
of the identical heavy quarks in the baryon arising from the
identical fermion exchange antisymmetry and
$(Q_1Q_2Q_3)_1^{(s,s_Z)}$ denotes  the color-singlet state
consisting of  three heavy quarks with  spin quantum number $s$ and
its third component $s_Z$. A heavy quark with a large mass moves in
the triply heavy baryon with a small velocity $v$ (in the baryon's
rest frame).
  $M(gg\ \ (q\bar{q})\ \
 \text{fusion})$ depending on $\overrightarrow{V}_1$ and $\overrightarrow{V}_2$ describes the hard
amplitude for producing the triple heavy quark pairs. The typical
momenta are of order  $m$ or higher. Thus in the leading order
(expansion in $v$) approximation, the $\overrightarrow{V}_1$,
$\overrightarrow{V}_2$ dependence in the matrix element $M(gg\ \
(q\bar{q})\ \ \text{fusion})$ can be neglected. The integration over
variables $\overrightarrow{V}_1$ and  $\overrightarrow{V}_2$ can be
carried out. It follows that from equation (\ref{total}):
\begin{eqnarray}\label{totals}
 &&A(gg\ \
(q\bar{q})\to\Omega_{Q_1Q_2Q_3}\bar{Q}_1\bar{Q}_2\bar{Q}_3)= \nonumber \\
 && \frac{\sqrt{2\overline{M}}}{\sqrt{8m_1\, m_2\,m_3}}
\frac{\Psi(0,0)}{\sqrt{d!}} M(gg \ (q\bar{q})\to
(Q_1Q_2Q_3)_1^{(s,s_Z)}\bar{Q}_1\bar{Q}_2\bar{Q}_3)\;,
\end{eqnarray}
where $\Psi(0,0)$ is value of the space wave function of the baryon
$\Omega_{Q_1Q_2Q_3}$ when the three quarks are all at the origin
which describes the long-distance effects that happen at energy
scale $mv$. In this way, the long-distance and short-distance
effects are separated at the amplitude level in the leading order
approximation. Our task now is to calculate the short-distance
amplitude $ M(gg \ \ (q\bar{q})\to
(Q_1Q_2Q_3)_1^{(s,s_Z)}\bar{Q}_1\bar{Q}_2\bar{Q}_3) $, which
describes triple heavy-quark pair production with the triple quarks
in a color-singlet and point-like $(Q_1Q_2Q_3)_1^{(s,s_Z)}$ state.
In calculating the amplitude, the momentum of each quark in the
$(Q_1Q_2Q_3)_1^{(s,s_Z)}$ state satisfies
$P_{Q_1}:P_{Q_2}:P_{Q_3}=m_1:m_2:m_3$.

In this paper, we consider only the triply heavy baryons production
of the $\Omega_{ccc}$, $\Omega^*_{ccb}$, and $\Omega_{ccb}$ states.
Since the production rates of other ones such as $\Omega_{bbb}$,
$\Omega^*_{cbb}$, and $\Omega_{cbb}$ are small, they are not
included in this paper.

\subsection{Feynman diagram generating}\label{sec}

\begin{figure}
\centering
\includegraphics[width=3cm]{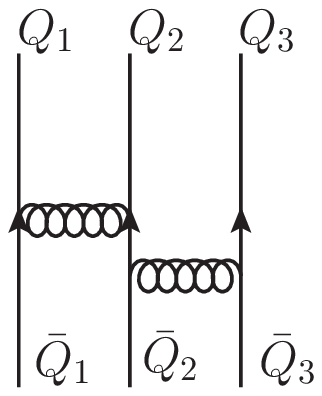}
\includegraphics[width=3cm]{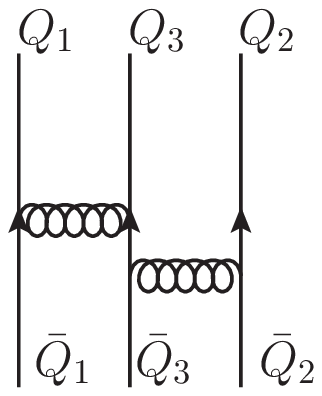}
\includegraphics[width=3cm]{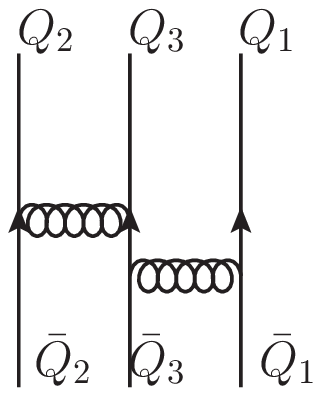}
\includegraphics[width=3cm]{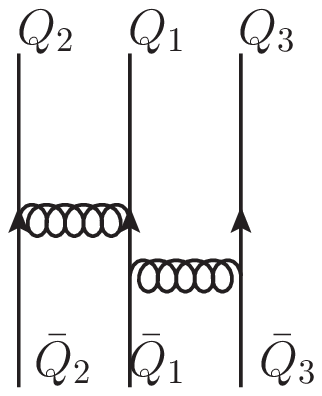}\\
\includegraphics[width=3cm]{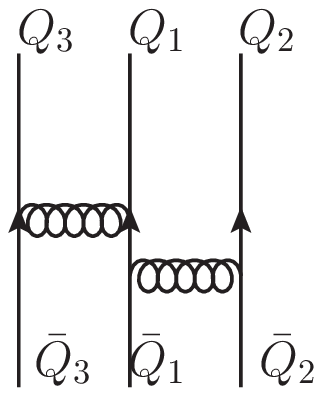}
\includegraphics[width=3cm]{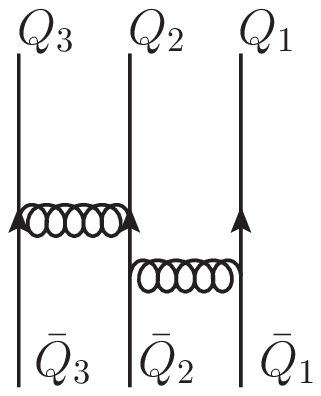}
\includegraphics[width=5cm]{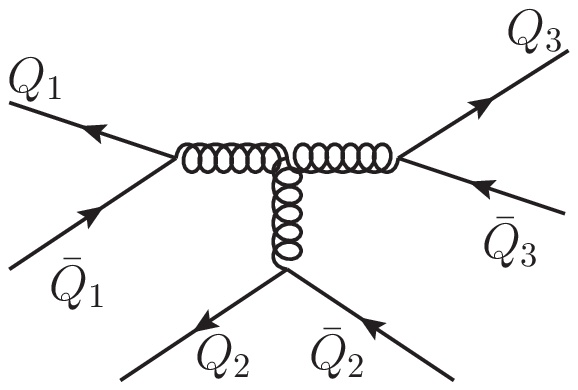}
\caption{Seven basic diagrams for generating the Feynman diagrams of
the triply heavy baryon production subprocess with different heavy
flavors. }\label{7dth}
\end{figure}

Since there are so many Feynman diagrams responsible for the
amplitude of the subprocess, to carry out the calculation, it is
crucial to generate the Feynman diagrams automatically. We now
propose a method to realize it by starting from the triply heavy
baryons production via $gg$ fusion subprocess.
 Notice that removing the external gluon lines and the
 corresponding interaction vertex for
$gg$ fusion subprocess, all Feynman diagrams reduce to seven basic
diagrams as shown in Fig.\ref{7dth} assuming three heavy quarks with
different heavy flavors. They are nothing but the diagrams
connecting three heavy quark lines in various ways. It means that
all  Feynman diagrams for $gg$ fusion subprocess can be obtained by
inserting two external gluon lines into all possible positions of
the 7 basic diagrams shown in Fig.\ref{7dth}.

For the production process of $\Omega_{ccc}$, the exchange
anti-symmetry of the identical fermions means that the number of the
basic Feynman diagrams is  $7 \times 3 ! = 42 $, while for the
production process of $\Omega_{ccb}$, it means that the number of
the basic Feynman diagrams is  $7 \times 2 ! = 14 $.

We now show how to generate all Feynman diagrams for the triple
heavy quark pairs production subprocess via $gg$ fusion. Notice that
every diagram contains at most single quadruple-gluon vertex. For
convenience, all diagrams are  divided into two classes in terms of
whether they contain the quadruple-gluon vertex or not.

We first generate Feynman diagrams without quadruple-gluon vertex by
inserting the external gluon lines in between the gluon or quark
lines of the basic diagrams one by one.  There are 3 internal and 6
external lines for each of the basic Feynman diagram.  The first
external gluon  can be inserted in between one of these 9 lines.
Then there are 4 internal and 7 external lines. And the second gluon
$g$ can be inserted in between one of these 11 lines. Therefore, for
the production process of the triply heavy baryon with three
different flavors, the number of the diagrams without
quadruple-gluon vertex is $ 7\times 9 \times 11=693 $.

We then generate the Feynman diagrams with one quadruple-gluon
vertex. They are generated either from the last diagram in
Fig.\ref{7dth}, where one gluon attaches to the triple-gluon vertex
while the other one inserts into all other possible lines, or from
all 7 basic diagrams with the double external gluon lines inserting
into the same point of any gluon line. These two ways generate $19$
and $15$ diagrams with quadruple-gluon vertex, respectively. Thus
totally one has $693 + 34 =727$ Feynman diagrams.

For the $\Omega_{ccb}$ ($\Omega^*_{ccb}$) and $\Omega_{ccc}$
production  via the $gg$ fusion subprocesses, there are $727\times 2
! = 1454$  and $727\times 3 ! = 4362$ Feynman diagrams,
respectively. In this way, we generate all Feynman diagrams
responsible for the tree level amplitude of the triply heavy baryon
production via the $gg$ fusion subprocesses.

In a similar way, we now can generate all Feynman diagrams for the
triple heavy quark pairs production via the quark-antiquark
annihilation subprocesses. In the production subprocess of
$q\bar{q}$ annihilating to three quark pairs with different flavors,
Feynman diagrams can be classified as three types in terms of the
number of gluon lines attached to the light quark lines. For the
first type diagrams with one gluon attached to the light quark line,
there are $7\times 9  = 63$ Feynman diagrams without quadruple-gluon
vertex and $1$ Feynman diagram with one quadruple-gluon vertex. For
the second type diagrams with double gluon lines attached to the
light quark lines, the Feynman diagrams can be generated by breaking
up one of the gluon lines in the 7 basic diagrams and attaching to
the light quark line. For this type, totally there are $2!\times
15=30$ Feynman diagrams. For the third type diagrams with triple
gluon lines attached to the light quark lines, all gluon lines
between heavy quarks are removed, and each of three gluon lines are
attached to each of three heavy quark lines. Thus for this type,
there are $3!\times 1=6$ diagrams. Adding  the three types of
Feynman diagrams together, totally  there are $64+30+6=100$ Feynman
diagrams for the process of $q\bar{q}$ annihilation to three heavy
quark pairs with different flavors. Thus for the $\Omega_{ccb}$
($\Omega^*_{ccb}$) and $\Omega_{ccc}$ production subprocesses via
the quark-antiquark annihilation, there are $100\times 2 ! = 200$
Feynman diagrams and $100\times 3 ! = 600$ ones, respectively.

\subsection{Color factors of the amplitude}

The large number of Feynman diagrams makes the calculation very
complicated. To carry out the calculations we first need to account
for the color factors. The amplitude can be classified into a set of
gauge invariant subsets in terms of variant independent color
factors. Denote $\xi_i=1,2,3$ , $\chi_i=1,2,3$ ($i=1,2,3$),  $a,\;
b=1,2,\cdots,8$, as the color indices of three heavy quarks, three
anti-heavy quarks, and two gluons, respectively. For the $gg$ fusion
production subprocess, there are totally 12 independent color
factors $C_{k}$ ($k =1,2,\cdots,12$) listed as follows:
\begin{eqnarray}\label{cof}
C_1&=&\frac{\varepsilon^{\xi_1\xi_2\xi_3}}{\sqrt{6}}
(\lambda^a\lambda^b)_{\xi_1\chi_2}\delta_{\xi_2\chi_1}
\delta_{\xi_3\chi_3}\;,\ \ \ \
C_2=\frac{\varepsilon^{\xi_1\xi_2\xi_3}}{\sqrt{6}}
(\lambda^b\lambda^a)_{\xi_1\chi_2}\delta_{\xi_2\chi_1}
\delta_{\xi_3\chi_3}\;,
\nonumber \\
C_3&=&\frac{\varepsilon^{\xi_1\xi_2\xi_3}}{\sqrt{6}}
(\lambda^a\lambda^b)_{\xi_1\chi_1}\delta_{\xi_2\chi_2}
\delta_{\xi_3\chi_3}\;, \ \ \ \
C_4=\frac{\varepsilon^{\xi_1\xi_2\xi_3}}{\sqrt{6}}
(\lambda^b\lambda^a)_{\xi_1\chi_1}\delta_{\xi_2\chi_2}
\delta_{\xi_3\chi_3}\;,
\nonumber \\
C_5&=&\frac{\varepsilon^{\xi_1\xi_2\xi_3}}{\sqrt{6}}
(\lambda^a\lambda^b)_{\xi_1\chi_3}\delta_{\xi_2\chi_1}
\delta_{\xi_3\chi_2}\;, \ \ \ \
C_6=\frac{\varepsilon^{\xi_1\xi_2\xi_3}}{\sqrt{6}}
(\lambda^b\lambda^a)_{\xi_1\chi_3}\delta_{\xi_2\chi_1}
\delta_{\xi_3\chi_2}\;,
\nonumber \\
C_7&=&\frac{\varepsilon^{\xi_1\xi_2\xi_3}}{\sqrt{6}} (\lambda^a
)_{\xi_1\chi_1}(\lambda^b)_{\xi_2\chi_2} \delta_{\xi_3\chi_3}\;,\ \
\ \ C_8=\frac{\varepsilon^{\xi_1\xi_2\xi_3}}
{\sqrt{6}}(\lambda^a)_{\xi_1\chi_2}(\lambda^b)_{\xi_2\chi_1}
\delta_{\xi_3\chi_3}\;,
\nonumber \\
C_9&=&\frac{\varepsilon^{\xi_1\xi_2\xi_3}}{\sqrt{6}}
(\lambda^a)_{\xi_1\chi_2}(\lambda^b)_{\xi_2\chi_3}
\delta_{\xi_3\chi_1}\;,\ \ \ \
C_{10}=\frac{\varepsilon^{\xi_1\xi_2\xi_3}}{\sqrt{6}}
(\lambda^a)_{\xi_1\chi_3}(\lambda^b)_{\xi_2\chi_2}
\delta_{\xi_3\chi_1}\;,
\nonumber \\
C_{11}&=&\frac{\varepsilon^{\xi_1\xi_2\xi_3}}{\sqrt{6}}
(\lambda^a)_{\xi_1\chi_3}(\lambda^b)_{\xi_2\chi_1}
\delta_{\xi_3\chi_2}\;,\ \ \ \
C_{12}=\frac{\varepsilon^{\xi_1\xi_2\xi_3}}{\sqrt{6}}
(\lambda^a)_{\xi_1\chi_1}(\lambda^b)_{\xi_2\chi_3}
\delta_{\xi_3\chi_2}\;.
\end{eqnarray}
The color factors given in Eq. (\ref{cof})  can be used as the
independent bases of the color wave function of the amplitude of the
$gg$ fusion subprocess. With these color factors, the amplitude of
each diagram can be expressed as the sum of the products of the
numerical values and those color factors. It yields:
\begin{eqnarray} \label{m-c-gg}
  \mathcal{M}&=&d!\sum_{k=1}^{12}C_k \,\Gamma_{k}\;,
\end{eqnarray}
where $\Gamma_{k}$ is the amplitude of the basic Feynman diagrams related to the color-factor $C_k$. 

Denote $\varsigma_1=1,2,3$ and $\varsigma_2=1,2,3$  the color
indices of $q$ and $\bar{q}$ for $q\bar{q}$ annihilation subprocess.
There are totally 4 independent color factors $D_{k}$ ($k$=1,2,3,4)
for the triple heavy quark pair production via $q\bar{q}$
annihilation subprocess listed as follows:
\begin{eqnarray}\label{cof2}
D_1=\frac{\varepsilon^{\xi_1\xi_2\xi_3}}{\sqrt{6}}
 \delta_{\varsigma_1\chi_1}\delta_{\varsigma_2\xi_1}
 \delta_{\xi_2\chi_2}
 \delta_{\xi_3\chi_3}\;,\nonumber
\\
D_2=\frac{\varepsilon^{\xi_1\xi_2\xi_3}}{\sqrt{6}}
\delta_{\varsigma_1\chi_2}\delta_{\varsigma_2\xi_1}
\delta_{\xi_2\chi_3}\delta_{\xi_3\chi_1}\;,
\nonumber \\
D_3=\frac{\varepsilon^{\xi_1\xi_2\xi_3}}{\sqrt{6}}
\delta_{\varsigma_1\chi_3}\delta_{\varsigma_2\xi_1}
\delta_{\xi_2\chi_1}\delta_{\xi_3\chi_2}\;, \nonumber \\
D_4=\frac{\varepsilon^{\xi_1\xi_2\xi_3}}{\sqrt{6}}
\delta_{\varsigma_1\varsigma_2}
\delta_{\xi_1\chi_1}\delta_{\xi_2\chi_2}\delta_{\xi_3\chi_3}\;.
\end{eqnarray}
The color factors given in Eq. (\ref{cof2}) can be used as the
independent bases of the color wave function of the amplitude of the
$q\bar{q}$ annihilation subprocess. With these color factors, the
amplitude of each diagram can be expressed as the sum of the
products of the numerical values and those color factors. It follows
that:
\begin{eqnarray} \label{m-c-qq}
  \mathcal{M}&=&d!\sum_{k=1}^{4}D_k \,\Gamma^{'}_{k}\;,
\end{eqnarray}
where $\Gamma^{'}_{k}$ is the amplitude of the basic Feynman diagrams related to the color-factor $D_k$. 

 For those diagrams without the quadruple-gluon vertex, the
amplitude of each diagram can be expressed as single product of a
numerical value and a color factor, each of which can be decomposed
as the sum of the color factors listed above. For those diagrams
with the quadruple-gluon vertex, the amplitude of each diagram can
be expressed as a sum of three products of numerical values and
color factors, again, each of which can be decomposed as the sum of
the color factors listed above.

\subsection{Simplification by identical Fermionic symmetry}

As discussed in Sec. \ref{sec}, the permutation of the identical
heavy quarks increases the number of Feynman diagrams by a factor
$3!$ and $2!$, for the production of color-singlet triple $c$ quark
state and double $c$ quarks plus single $b$ quark one, respectively.
Consequently, there are 4362 and 1454 Feynman diagrams that need to
be calculated for the amplitudes of $\Omega_{ccc}$ and
$\Omega_{ccb}$ ($\Omega^*_{ccb}$), respectively, via $gg$ fusion
subprocesses. For the quark-antiquark annihilation subprocesses, the
number of the Feynman diagrams are 200 and 600 for the production
amplitudes of the $\Omega_{ccc}$ and the $\Omega_{ccb}$
($\Omega^{*}_{ccb}$), respectively. Notice that the momenta of the
identical heavy quarks in triply heavy baryons can be taken to be
equal in calculating the amplitudes. This feature together with the
identical fermion exchanging symmetry can be used to simplify the
calculation. In fact, the contributions to the amplitude from those
increased diagrams by each permutation of the identical heavy quarks
are equal to the original diagrams. Thus one needs to calculate only
those 727 diagrams generated via 7 basic diagrams for the $gg$
fusion subprocesses, and those 100 diagrams for the $q\bar{q}$
annihilation subprocesses. Thus the calculations of the amplitudes
can be simplified significantly. We emphasize here that it is
available only for the amplitude of the leading order matrix
elements.

\subsection{Automatic amplitude calculation}

With the simplification in the last subsection, the number of the
Feynman diagrams that need to be calculated is reduced to 727 and
100 for the production of the triply heavy baryons via the
gluon-gluon fusion subprocess and the quark-antiquark annihilation
subprocess, respectively. These are still very large numbers for
doing numerical calculations. It would be quite lengthy and very
tedious to  code for numerical calculations if they are evaluated
and expressed by the conventional method. Instead we develop a
method to calculate Feynman diagrams closely related to the method
for generating Feynman diagrams described in Sec. 2.2, incorporating
a direct, numerical amplitude calculation method given in Ref.
\cite{B-Ph} as described as follows.

Given the quantum numbers and momenta of the initial and final
states in the amplitude, all momenta of the internal lines are fixed
in a specific tree level Feynman diagram. The propagators and
vertices in the Feynman diagram are described by certain matrix or
tensor while the spinor wavefunctions of the initial state and final
state are described by the line and raw vectors,
respectively.\footnote{In Appendix \ref{sec-3}, we list the
4-dimension vectors describing the spinors of the quark and
antiquark, and the polarization vectors of the gluons. } Then the
calculation of the amplitude of the corresponding Feynman diagram is
nothing but the product of the matrices to a numerical number
against Fermion lines. The method turns out to be efficient in
numerical calculation of tree level amplitude. However, for
subprocesses with too many Feynman diagrams like the triply heavy
baryons production one, it is still difficult to write down the
amplitude of all diagrams one by one.

To overcome this issue, we develop an automatic calculation method,
closely following the method for generating Feynman diagrams
described in Sec. \ref{sec}. Notice that the Feynman diagrams differ
from each other by the positions and the types of propagators and
vertices. Each type is assigned to a fix number. A Feynman diagram
is then represented by a table consisting of a set of ordered
numbers, each of which corresponds to the type of the propagator or
vertex in the diagram.  With the quantum numbers and the momenta of
each propagator or vertex, the Feynman diagram can be calculated by
the multiplication of the matrices or tensors controlled by the
number in the table. 
 In this way,
the program code for calculating the amplitudes can be written in a
compact form and contributions from all diagrams can be calculated
conveniently. In numerical simulation of each event, we assign the
momenta, helicities of the light quarks, the heavy antiquarks, and
the heavy baryon, and the polarization vectors of gluons randomly,
assuming that they meet a set of conservation laws. We then use
Vegas to simulate the events and carry out the phase space
integrals.

To ensure the correctness of the  program code, we examine various
gauge invariance of the amplitude by Ward-Takahashi identities of
the $gg$ fusion subprocess. Namely, substituting one of two
polarization vectors of gluons into its momentum, the total
amplitude vanishes. It follows from Eq. (\ref{m-c-gg}) that  each
amplitude $\Gamma_k$ ($k$=1,2,...,12) vanishes under the
substitution. To have additional test, in a similar way, we can also
examine the triply heavy baryons production processes via double
photons and $\gamma g$ fusion subprocess, in which the amplitudes
are expressed as the sum of different gauge independent subsets.
Similarly, we can also have various test for the  program code for
the subprocess of the $q\bar{q}$ annihilation.

\subsection{The cross sections of the subprocess}
We have calculated the amplitude  of the subprocesses $gg\ \
(q\bar{q})$ to triple heavy quark pairs above. When the amplitude is
expressed as Eq. (\ref{m-c-gg}), using the formula given  in
Appendix \ref{sec-2}, the squared amplitude for the $gg$ fusion
subprocess reads:
\begin{eqnarray}\label{sq-am-1}
\sum_{a,b,\chi_i}|\mathcal{M}|^2&= &(d!)^2\bigg(\frac{64}{9}\sum^{12}_{k=1}|\Gamma_{k}|^2+2   Re[ -\frac{8}{9}\sum_{k=1,3,5} \Gamma_k  \Gamma^*_{k+1}  \nonumber \\
&+&\frac{20}{9}(\Gamma_1  \Gamma^*_3+\Gamma_1  \Gamma^*_5+\Gamma_2
\Gamma^*_4 +\Gamma_2  \Gamma^*_6-\Gamma_3  \Gamma^*_5-\Gamma_4
\Gamma^*_6 +\Gamma_7  \Gamma^*_8+\Gamma_9  \Gamma^*_{10}\nonumber \\
&+&\Gamma_{11} \Gamma^*_{12}) +\frac{32}{9}(\Gamma_1
\Gamma^*_7-\Gamma_1 \Gamma^*_{10}-\Gamma_2 \Gamma^*_8 +\Gamma_2
\Gamma^*_9+\Gamma_3  \Gamma^*_8-\Gamma_3
\Gamma^*_{11} -\Gamma_4  \Gamma^*_{7}\nonumber \\
&+&\Gamma_4 \Gamma^*_{12}-\Gamma_5 \Gamma^*_{9} +\Gamma_{5}
\Gamma^*_{12}+\Gamma _{6}  \Gamma^*_{10}-\Gamma_{6} \Gamma^*_{11}
+\Gamma_{7}  \Gamma^*_{10}+\Gamma_{7}  \Gamma^*_{12} +\Gamma_{8}
\Gamma^*_9 \nonumber \\
&+&\Gamma_{8}  \Gamma^*_{11}+\Gamma_{9} \Gamma^*_{12} +\Gamma_{10}
\Gamma^*_{11}) -\frac{16}{9}(\Gamma_{1}  \Gamma^*_{4}+\Gamma_{1}
\Gamma^*_{6}+\Gamma_{1}  \Gamma^*_{11} -\Gamma _{1}
\Gamma^*_{12} \nonumber \\
&+&\Gamma_{2}  \Gamma^*_{3}+\Gamma_{2}  \Gamma^*_{5} +\Gamma_{2}
\Gamma^*_{11}-\Gamma_{2} \Gamma^*_{12}-\Gamma_{3} \Gamma^*_{6}
 -\Gamma_{3}  \Gamma^*_{9}+\Gamma_{3}
\Gamma^*_{10}-\Gamma_{4} \Gamma^*_{5} \nonumber \\
&-&\Gamma_{4} \Gamma^*_{9}+\Gamma_{4}  \Gamma^*_{10}- \Gamma_{5}
\Gamma^*_{7} +\Gamma_{5}  \Gamma^*_{8}-\Gamma_{6}  \Gamma^*_{7}
+\Gamma_{6}  \Gamma^*_{8} -\Gamma_{7}  \Gamma^*_{9}-\Gamma_{7}
\Gamma^*_{11}\nonumber \\
&-&\Gamma_{8} \Gamma^*_{10} -\Gamma_{8} \Gamma^*_{12}-\Gamma_{9}
\Gamma^*_{11}-\Gamma_{10}  \Gamma^*_{12}) +\frac{4}{9}(\Gamma_{1}
\Gamma^*_{8}-\Gamma_{1} \Gamma^*_{9}-\Gamma_{2}  \Gamma^*_{7}
+\Gamma _{2}  \Gamma^*_{10} \nonumber \\
&+&\Gamma_{3} \Gamma^*_{7}-\Gamma_{3} \Gamma^*_{12} -\Gamma_{4}
\Gamma^*_{8}+\Gamma_{4} \Gamma^*_{11}-\Gamma_{5} \Gamma^*_{10}
+\Gamma_{5} \Gamma^*_{11}+\Gamma_{6}
 \Gamma^*_{9}-\Gamma_{6} \Gamma^*_{12})]\bigg)\;, \nonumber \\
\end{eqnarray}
where, $Re$ means taking the real part of the number, and
$\sum_{a,b,\chi_i}$ means the summation of the color indices of the
initial gluons and terminal heavy antiquarks.

Similarly, with the amplitude given in Eq. (\ref{m-c-qq}), the
squared amplitude for the $q\bar{q}$ annihilation subprocess reads:
\begin{eqnarray}\label{sq-am-qq}
\sum_{\zeta_1,\zeta_2,\chi_i}|\mathcal{M}|^2&=&(d!)^2\bigg(3\sum^4_{i=1}\Gamma_i\Gamma^*_i
-2\cdot Re(\Gamma^{'}_1 \Gamma^{'*}_2+\Gamma^{'}_1
\Gamma^{'*}_3+\Gamma^{'}_2 \Gamma^{'*}_3)\nonumber \\
&+&2\cdot Re(\Gamma^{'}_1 \Gamma^{'*}_4+\Gamma^{'}_2
\Gamma^{'*}_4+\Gamma^{'}_3 \Gamma^{'*}_4)\bigg)\;.
\end{eqnarray}

 The differential cross section of the subprocess is then given by:
\begin{eqnarray} \label{dif-cross}
  && d\hat{\sigma}(gg \ \ (q\bar{q})\to
  \Omega_{Q_1Q_2Q_3}\bar{Q}_1\bar{Q}_2\bar{Q}_3
  ) \nonumber \\=&& \frac{1}{d!}d\Pi_4 \;
  \frac{1}{(2\pi)^8\,2\hat{s}} \;
  \frac{8}{N} \;
  \sum_{s_Z}\sum_{a,b,\varsigma_1,\varsigma_2,\chi_i}
  |A(gg\ \ (q\bar{q})\to
  \Omega_{Q_1Q_2Q_3}\bar{Q}_1\bar{Q}_2\bar{Q}_3)|^2\;,
 \end{eqnarray}
  with,
\begin{eqnarray}
  d\Pi_4&=&\frac{d^3P}{2E}\,
   \frac{d^3q_1}{2E_{q_1}}\,
   \frac{d^3q_2}{2E_{q_2}}\,
   \frac{d^3q_3}{2E_{q_3}}\;
   \delta^4(k_1+k_2-P-q_1-q_2-q_3)\;, \nonumber
\end{eqnarray}
where $\hat{s}$ is the subprocess c.m. energy squared, and the
denominator $N$ in $\frac{8}{N}$ is 64 and 9 for the $gg$ fusion
subprocess and the $q\bar{q}$ annihilation one, respectively, and
the numerator 8 in $\frac{8}{N}$  means the summation of the spins
of the  heavy antiquarks in the final states,  which are given
randomly in the code. The spins of the initial gluons (light quark
and antiquark) are given randomly in the code too.

\section{Numerical results and conclusions}

Incorporating the parton model with the cross sections of the
subprocesses, the cross sections of the process $pp\to
\Omega_{Q_1Q_2Q_3}\bar{Q}_1\bar{Q}_2\bar{Q}_3+X$ in $pp$ collisions
can be calculated. It follows that:
\begin{eqnarray} \label{cross}
  \sigma&=&\int dx_1dx_2f_{g_1}(x_1,\mu_F)
  f_{g_2}(x_2,\mu_F)\;\int
  d\hat{\sigma}(gg\to
  \Omega_{Q_1Q_2Q_3}\bar{Q}_1\bar{Q}_2\bar{Q}_3, \mu_F)
\nonumber \\
  &+&\sum_{q}\int dx_1dx_2f_{q}(x_1,\mu_F)
  f_{\bar{q}}(x_2,\mu_F)\;
  \int d\hat{\sigma}(q\bar{q}\to \Omega_{Q_1Q_2Q_3}
  \bar{Q}_1\bar{Q}_2\bar{Q}_3,\mu_F)\;
  ,
\end{eqnarray}
where $f_g(x,\mu_F)$ and $f_q(x,\mu_F)$ ($f_{\bar{q}}(x,\mu_F)$) are
the parton distribution functions of the gluon and quark (antiquark)
in the proton, respectively, $\mu_F$ is the factorization energy
scale, $\sum_q$ means summation of all light flavors of $q$. It
happens that the predicted cross sections in the leading order are
just the same for the production of the baryons consisting of
identical fermions and non-identical ones. It is no longer valid
beyond the leading order.

Given the  mass parameters $m_c=1.5 $ GeV and $m_b=4.9$ GeV, the
values of  $|\Psi(0,0)|$'s can be evaluated by the formula given in
Ref. \cite{xm4}. They are  0.0781 GeV$^3$, 0.0864 GeV$^3$, for the
$\Omega_{ccc}$ and $\Omega_{ccb}$ ($\Omega^*_{ccb}$) states,
respectively.

Our numerical results show that the contributions to the total cross
sections from the  $gg$ fusion subprocesses dominate as expected. In
fact, contributions from the  $gg$ fusion subprocesses are 2$\sim$3
orders of magnitude higher than that from the $q\bar{q}$
annihilation subprocesses. Thus contributions to the cross sections
from the $q-\bar{q}$ annihilation subprocesses are negligible.

To carry our numerical calculations we take  Cteq6l \cite{distri}
parton distribution functions.  Some uncertainties arise from the
ambiguities in choosing the factorization energy scale $\mu_F$. For
comparison, here we take two different values of $\mu_F$, i.e.,
$\mu_F$=$\mu_R/2$ and $\mu_F$=$\mu_R$, where
$\mu_R^2$=$P_T^2$+$\overline{M}^2$.

We calculate the  production cross sections of the baryons
$\Omega_{ccc}$, $\Omega_{ccb}^*$ and $\Omega_{ccb}$ at LHC with
total energy $\sqrt{S}=7 $ TeV and $\sqrt{S}= 14$ TeV and list in
Table \ref{cross:7tev} and Table \ref{cross:14tev}  with various
$P_T$ and pseudo-rapidity cuts, respectively. From the tables, we
see that the predicted numerical results differ by around factor 2
with two different factorization scales $\mu_F$=$\mu_R/2$ and
$\mu_F$=$\mu_R$. These uncertainties are expected to be reduced by
including the next leading calculations which will be extremely
difficult work.

\begin{table*}
\caption{Predicted  cross sections (in unit $nb$) of the triply
heavy baryon production at LHC with $\sqrt{S}=7$ TeV. Three typical
$P_T$ cuts are adopted. Pseudo-rapidity cuts $|\eta|< 2.5$ for CMS
and ATLAS, and $1.9<\eta< 4.9$ for LHCb are taken. }
\label{cross:7tev}
\begin{tabular}{|c|c|c|c|c|c|}
\hline\hline
-&-& \multicolumn{2}{|c|}{LHC (CMS, ATLAS) }& \multicolumn{2}{|c|}{~~LHCb~~} \\
\hline
 -&\backslashbox{$P_{Tcut}$} {$\eta_{cut}$}& \multicolumn{2}{|c|}{$|\eta|< 2.5$ }&\multicolumn{2}{|c|}{ $1.9<\eta< 4.9$}\\
\hline $\mu_F$&-&$\mu_R$&$\mu_R/2$&$\mu_R$&$\mu_R/2$ \\
\hline$\Omega_{ccc}$&0GeV&0.0604 &0.132 &0.0329 &0.0724 \\
-&5GeV&0.00599 &0.0140 &0.00163&0.00391\\
-&10GeV&2.6 E-4&6.3E-4&4.8E-5&1.21E-4\\
\hline
$\Omega_{ccb}^*$&0GeV&0.00151&0.00351&7.24E-4&0.00172\\
-&5GeV&6.49E-4&0.00152&1.89E-4&4.54E-4\\
-&10GeV&9.62E-5&2.26E-4&1.95E-5&4.67E-5\\
\hline
$\Omega_{ccb}$&0GeV&4.89E-4&0.00114&2.15E-4&5.09E-4\\
-&5GeV&2.43E-4&5.67E-4&6.86E-5&1.65E-4\\
-&10GeV&4.49E-5&1.05E-4&0.894E-5&2.13E-5\\
\hline
\end{tabular}
\end{table*}

\begin{table*}
\caption{ Predicted  cross sections (in unit $nb$) of the triply
heavy baryon production at LHC with $\sqrt{S}=14$ TeV. Three typical
$P_T$ cuts are adopted. Pseudo-rapidity cuts $|\eta|< 2.5$ for CMS
and ATLAS, and $1.9<\eta< 4.9$ for LHCb are taken. } \vspace{2mm}
\begin{tabular}{|c|c|c|c|c|c|}
\hline\hline
-&-& \multicolumn{2}{|c|}{LHC (CMS, ATLAS) }& \multicolumn{2}{|c|}{~~LHCb~~} \\
\hline
 -&\backslashbox{$P_{Tcut}$} {$\eta_{cut}$}& \multicolumn{2}{|c|}{$|\eta|< 2.5$ }&\multicolumn{2}{|c|}{ $1.9<\eta< 4.9$}\\
\hline $\mu_F$&-&$\mu_R$&$\mu_R/2$&$\mu_R$&$\mu_R/2$ \\
\hline$\Omega_{ccc}$&0GeV&0.113&0.216&0.0684&0.135\\
-&5GeV&0.0123&0.0258&0.00412&0.00906\\
-&10GeV&0.000625&0.00136&0.000145&0.000349\\
\hline
$\Omega_{ccb}^*$&0GeV&0.00320&0.00677&0.00175&0.00378\\
-&5GeV&0.00143&0.00307&0.000521&0.00114\\
-&10GeV&2.34E-4&5.03E-4&0.625E-4&1.38E-4\\
\hline
$\Omega_{ccb}$&0GeV&0.00105&0.00222&0.000527&0.00115\\
-&5GeV&0.000544&0.00117&0.000190&0.000419\\
-&10GeV&0.000109&0.000236&0.289E-4&0.639E-4\\
\hline
\end{tabular}\label{cross:14tev}
\end{table*}

From Table \ref{cross:7tev}, we see that for $10 fb^{-1}$ integrated
luminosity running at $7$ TeV, for the production of $\Omega_{ccc}$,
around $(0.6-1.4)\times 10^{5}$ events at CMS and ATLAS can be
accumulated with kinematic cuts $P_T > 5$ GeV and $|\eta| < 2.5$,
while this number is around $(2-4)\times 10^{4}$ with kinematic cuts
$P_T > 5$ GeV and $1.9< \eta < 4.9 $ at LHCb.  For the production of
$\Omega_{ccb}$, both $\Omega_{ccb}$ and $\Omega^*_{ccb}$ needs to
add together since the higher mass state will decay into the lower
mass state by emitting a photon. After doing this,  we see that for
$10 fb^{-1}$ integrated luminosity running at $7$ TeV, for the
production of $\Omega_{ccb}$, around $(1-2)\times 10^{4}$ events at
CMS and ATLAS can be accumulated with kinematic cuts $P_T > 5$ GeV
and $|\eta| < 2.5$, while this number is around $(2-6)\times 10^{3}$
with kinematic cuts $P_T > 5$ GeV and $1.9< \eta < 4.9 $  at LHCb.

From Table \ref{cross:14tev}, we see that for $1000 fb^{-1}$
integrated luminosity running at $14$ TeV, for the production of
$\Omega_{ccc}$, around $(0.6-1.4)\times 10^{6}$ events at CMS and
ATLAS can be accumulated with kinematic cuts $P_T > 10$ GeV and
$|\eta| < 2.5$, while this number is around $(4-9)\times 10^{6}$
with kinematic cuts $P_T > 5$ GeV and $1.9< \eta < 4.9$ at LHCb. For
the production of $\Omega_{ccb}$, again by adding both
$\Omega_{ccb}$ and $\Omega^*_{ccb}$ events together,  we see that
for $1000 fb^{-1}$ integrated luminosity running at $14$ TeV, for
the production of $\Omega_{ccb}$, around $(3-7)\times 10^{5}$ events
at CMS and ATLAS can be accumulated with kinematic cuts $P_T > 10$
GeV and $|\eta| < 2.5$ while this number is around $(0.7-1.5)\times
10^{6}$ with kinematic cuts $P_T > 5$ GeV and $1.9< \eta < 4.9 $ at
LHCb.

Here we calculate only the contributions from the ground states.
Excited states also can decay into the ground states. Including
those contributions, the real production rates should be $2-3$ times
higher than ones listed in Table \ref{cross:7tev} and Table
\ref{cross:14tev}. Moreover, the cross sections for the anti-triply
heavy baryons are exactly the same. This will double the accumulated
triply heavy baryon events.
 From the tables, we see that the cross sections for LHC
running at $14$ TeV is about  2 times larger than that for LHC
running at $7$ TeV.

\begin{figure}
\centering
\includegraphics[width=9cm]{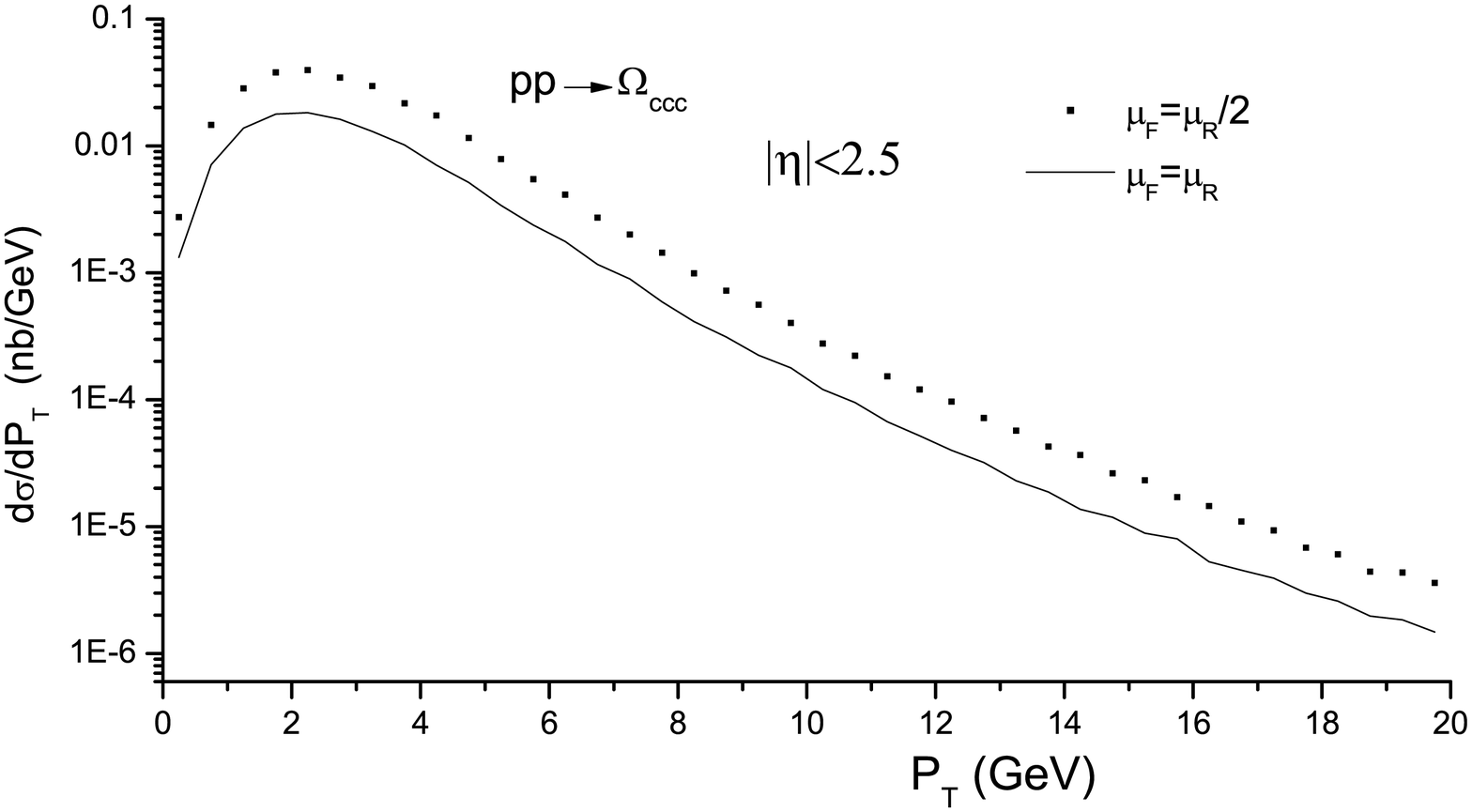}
\includegraphics[width=9cm]{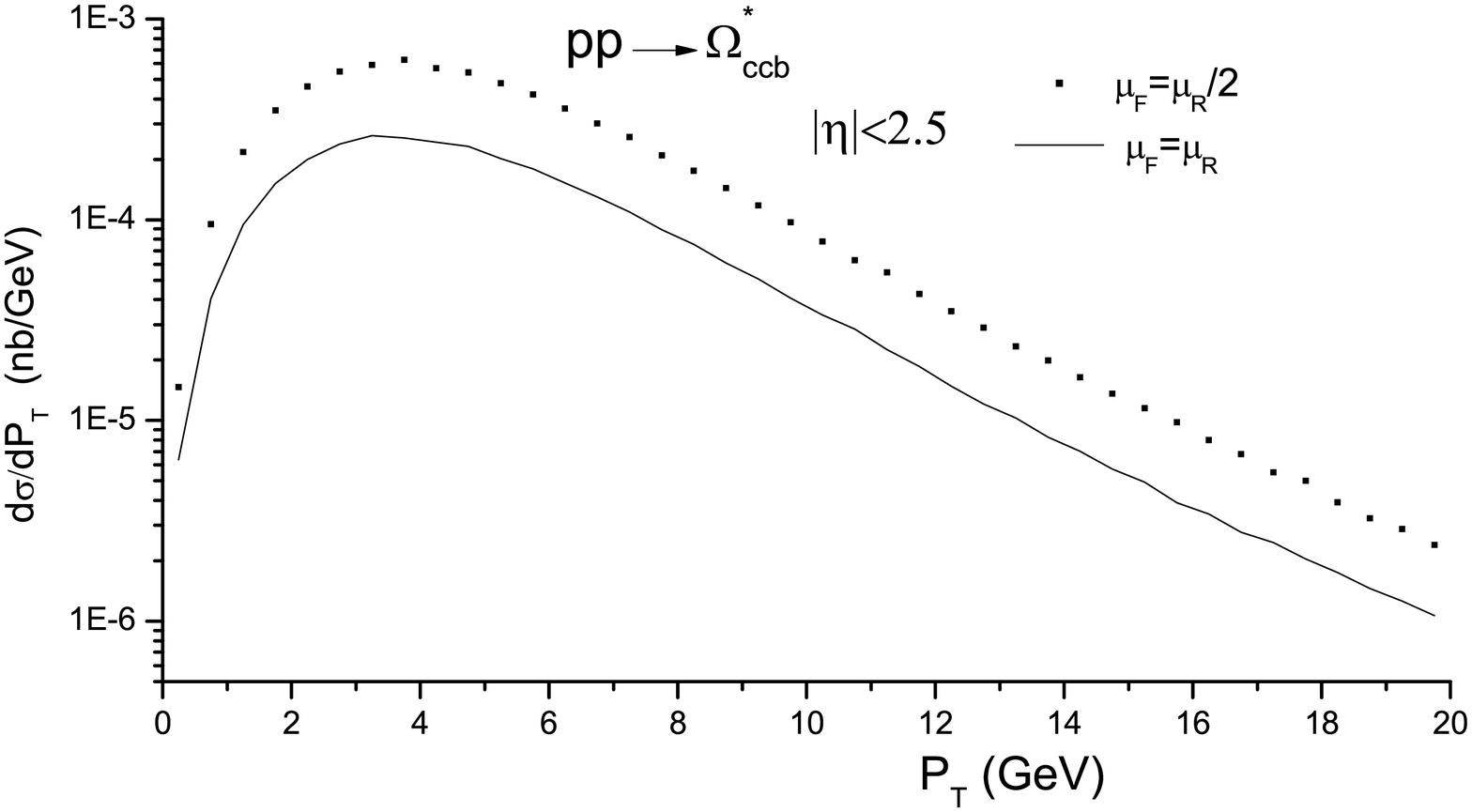}
\includegraphics[width=9cm]{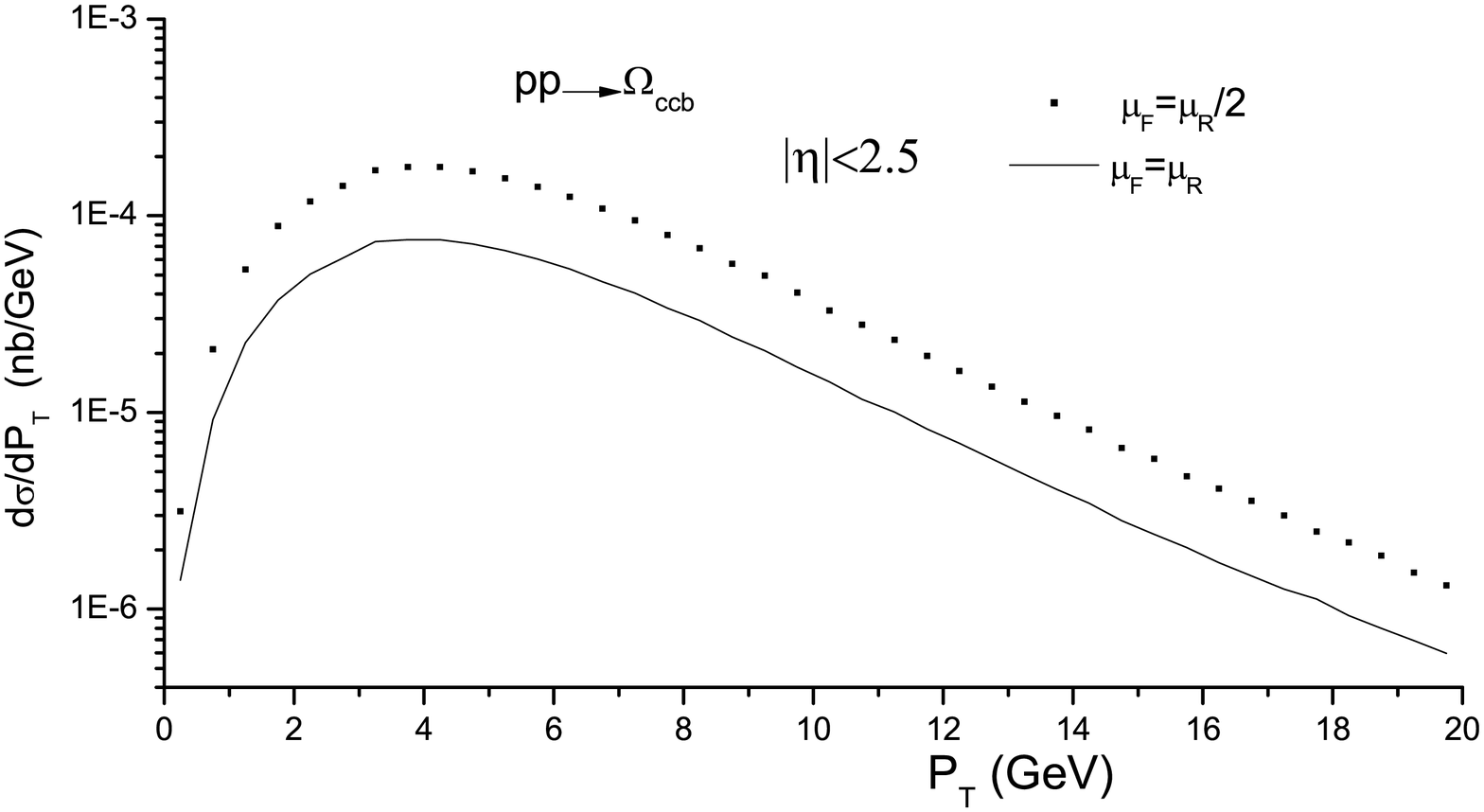}
\caption{The predicted $P_T$-distributions of the triply heavy
baryon production in CMS and ATLAS, with $\sqrt{S}$=7
TeV.}\label{ptcms:7tev}
\end{figure}

\begin{figure}
\centering
\includegraphics[width=9cm]{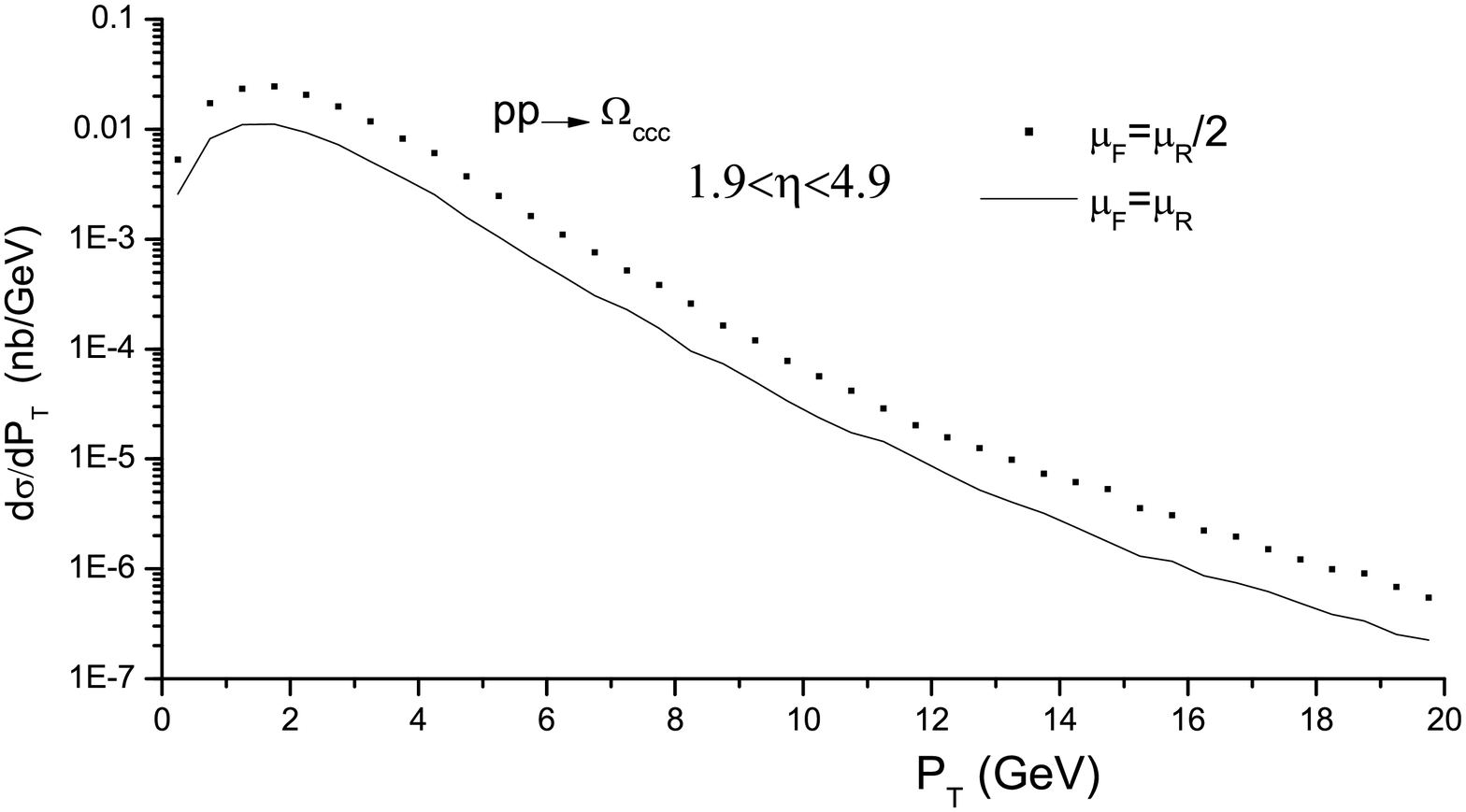}
\includegraphics[width=9cm]{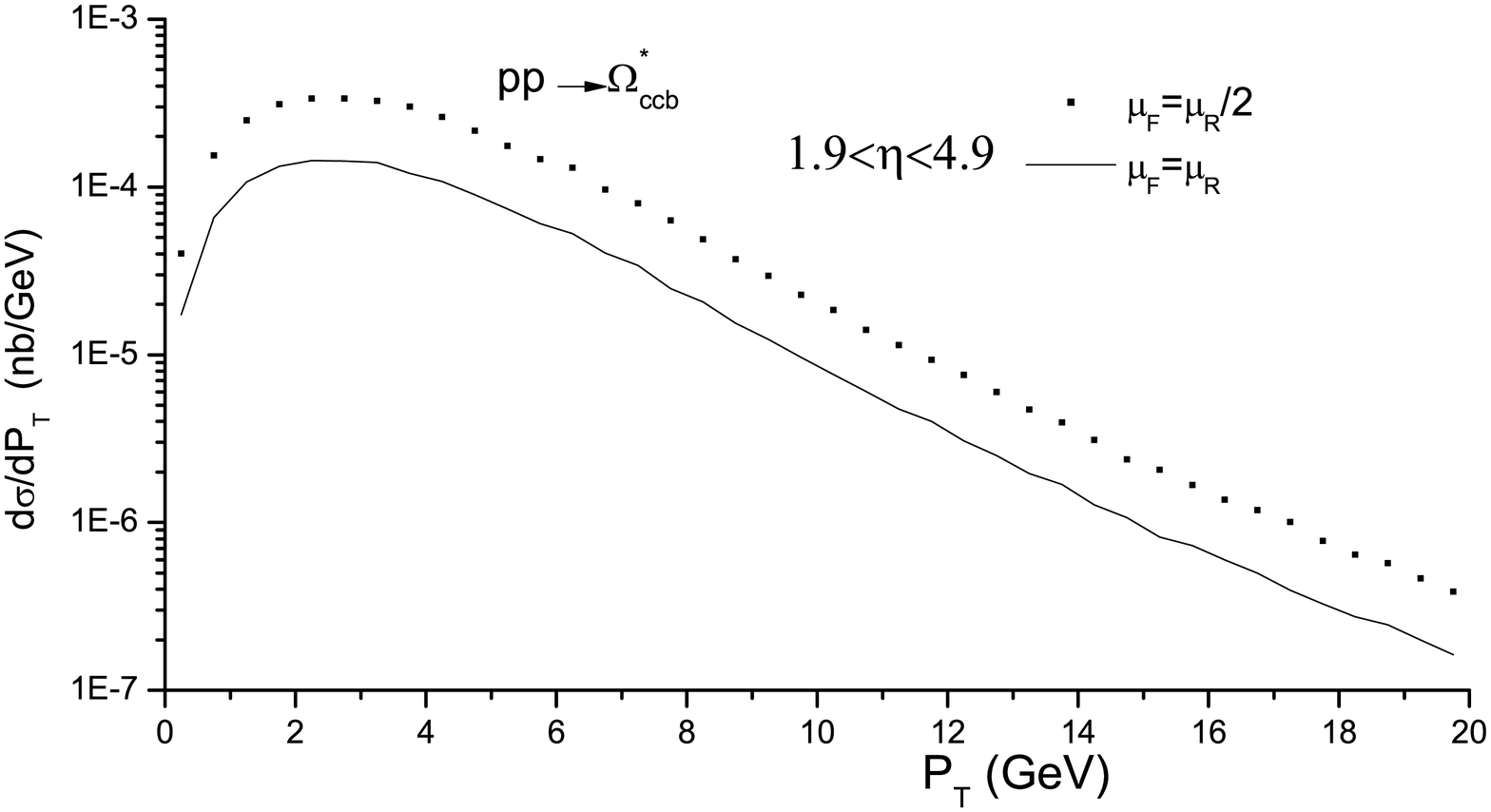}
\includegraphics[width=9cm]{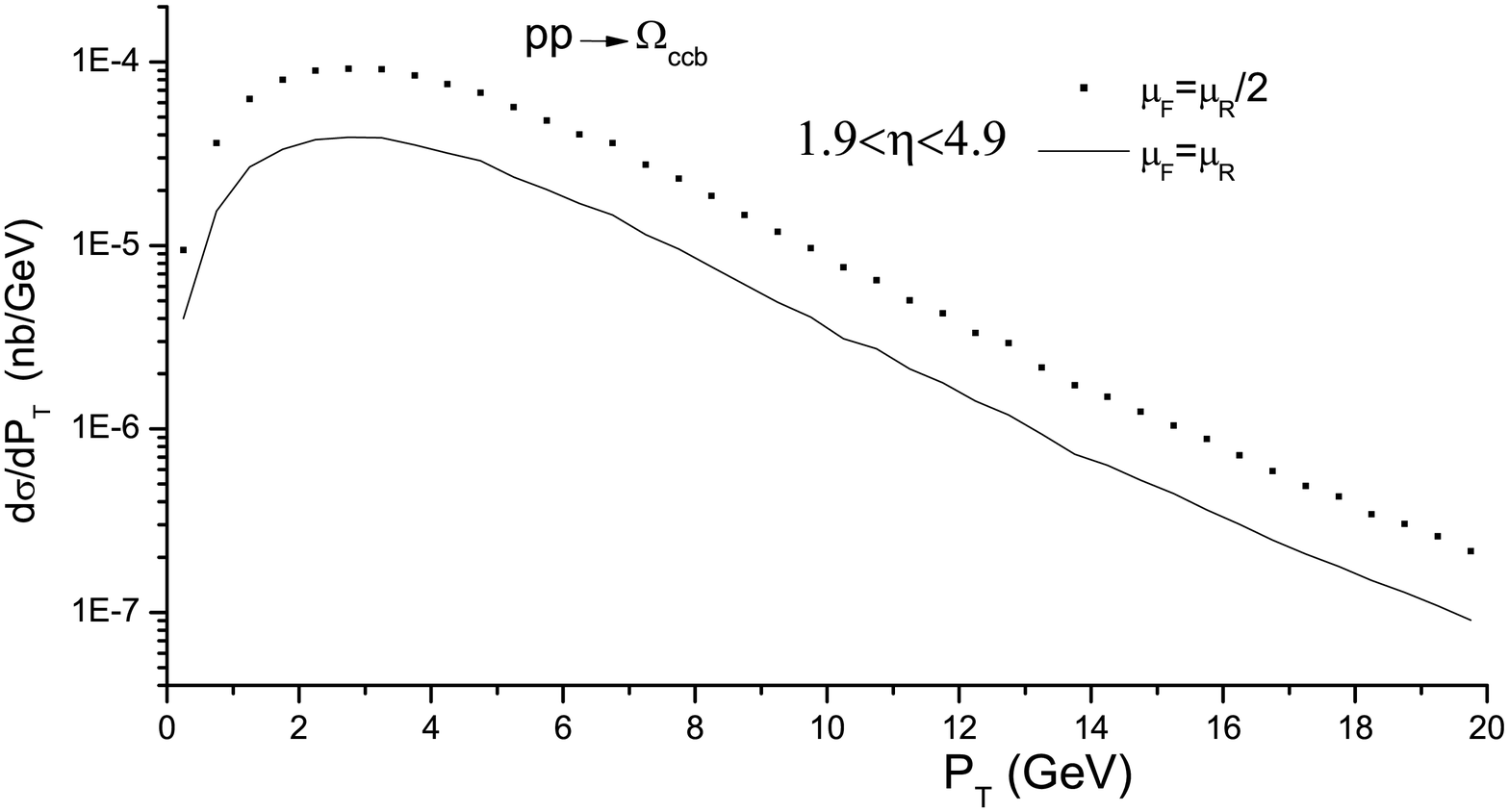}
\caption{The predicted $P_T$-distributions of the triply heavy
baryon production in LHCb  with $\sqrt{S}$=7 TeV.
}\label{ptlhc:7tev}
\end{figure}

\begin{figure}
\centering
\includegraphics[width=9cm]{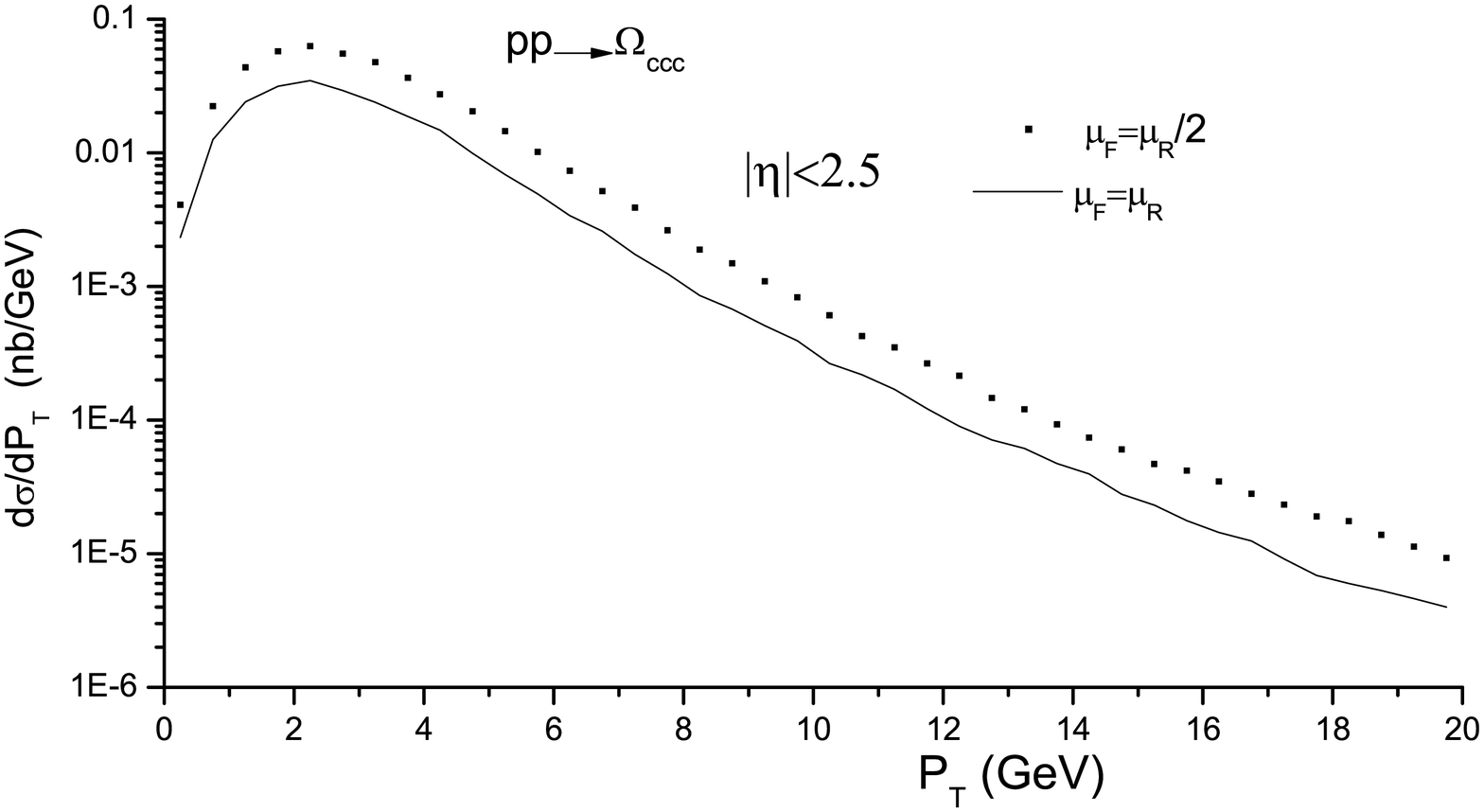}
\includegraphics[width=9cm]{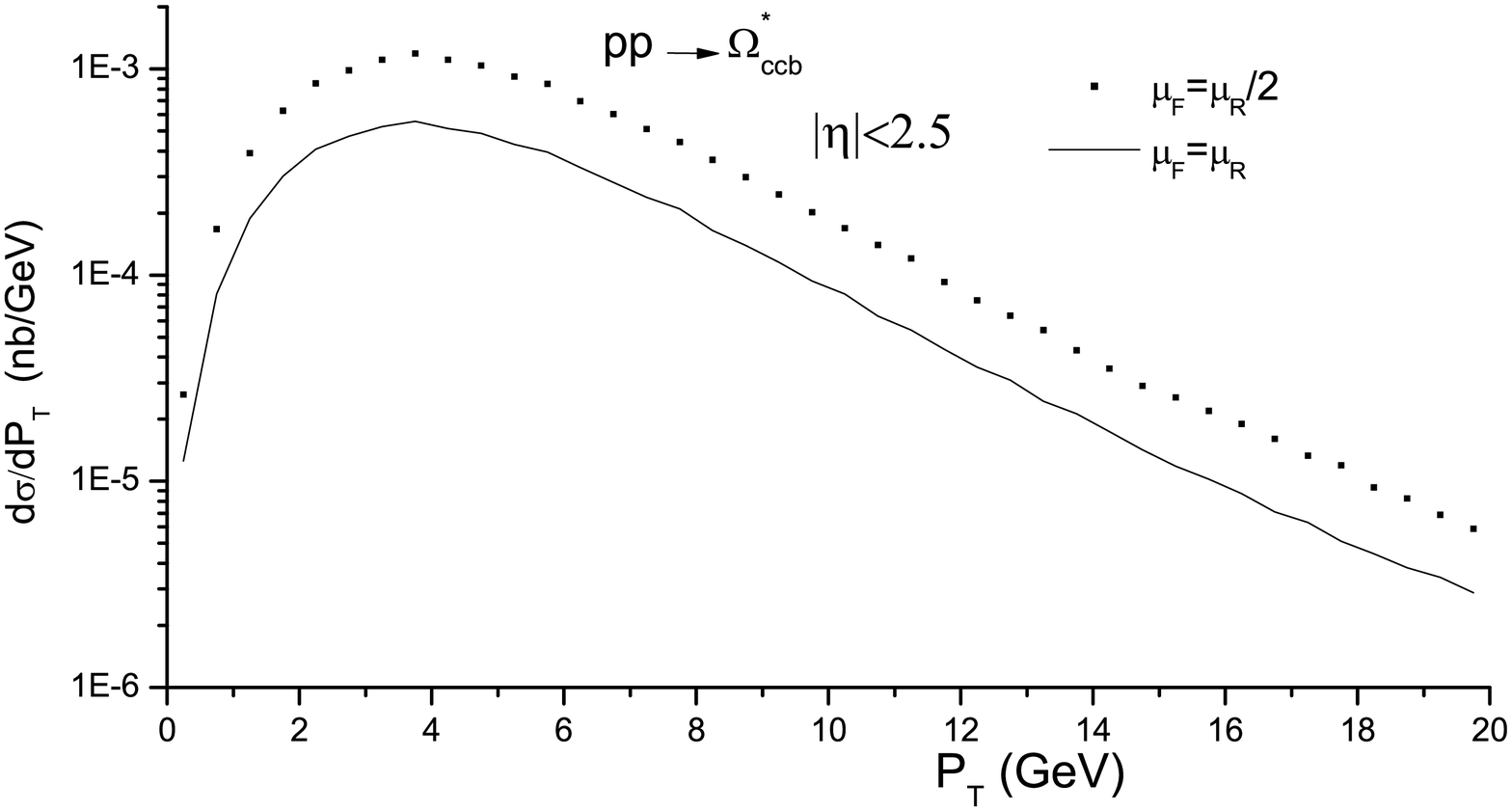}
\includegraphics[width=9cm]{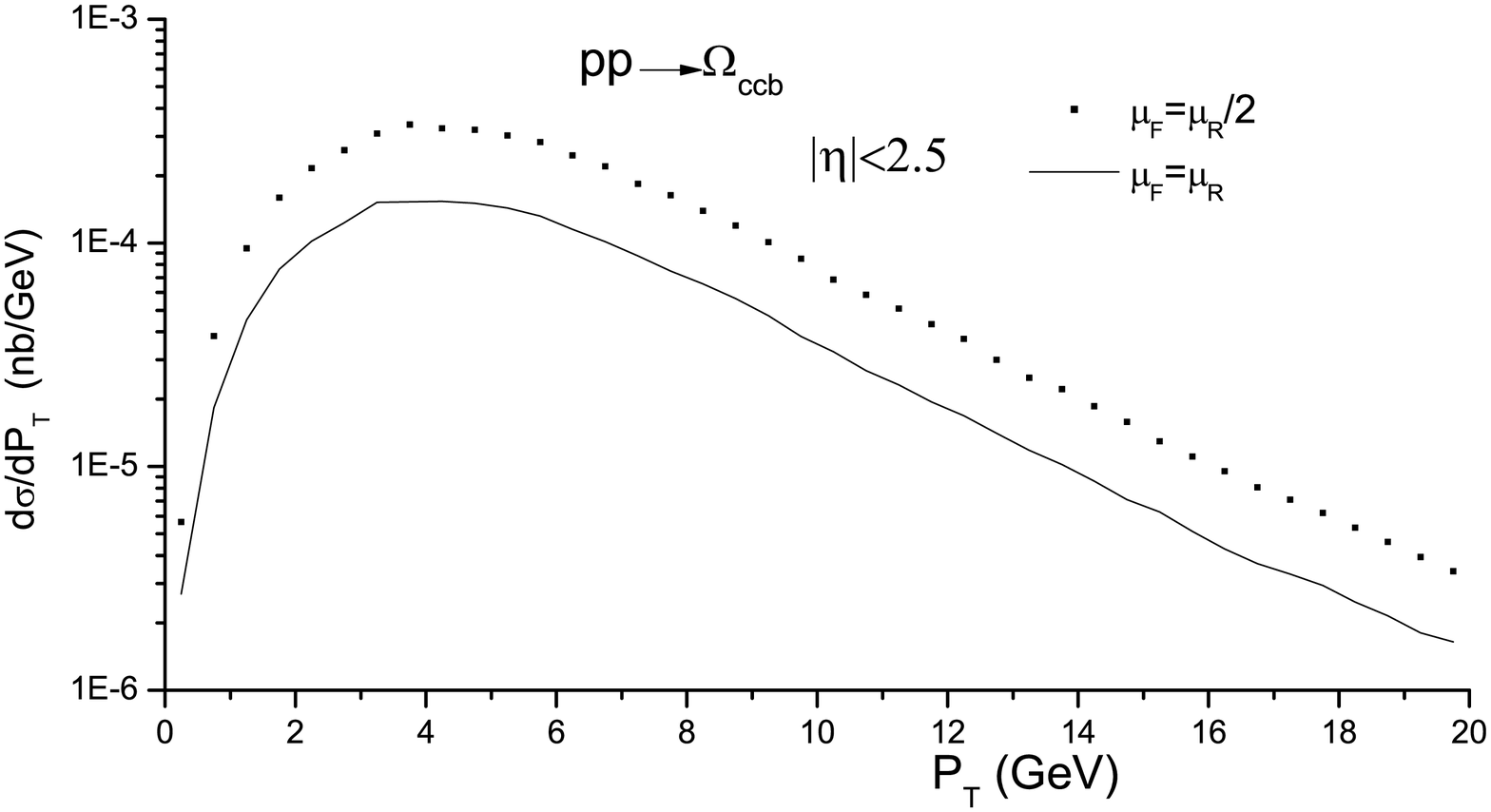}
\caption{The predicted $P_T$-distributions of the triply heavy
baryon production in CMS and ATLAS, with $\sqrt{S}$=14
TeV.}\label{ptcms}
\end{figure}

\begin{figure}
\centering
\includegraphics[width=9cm]{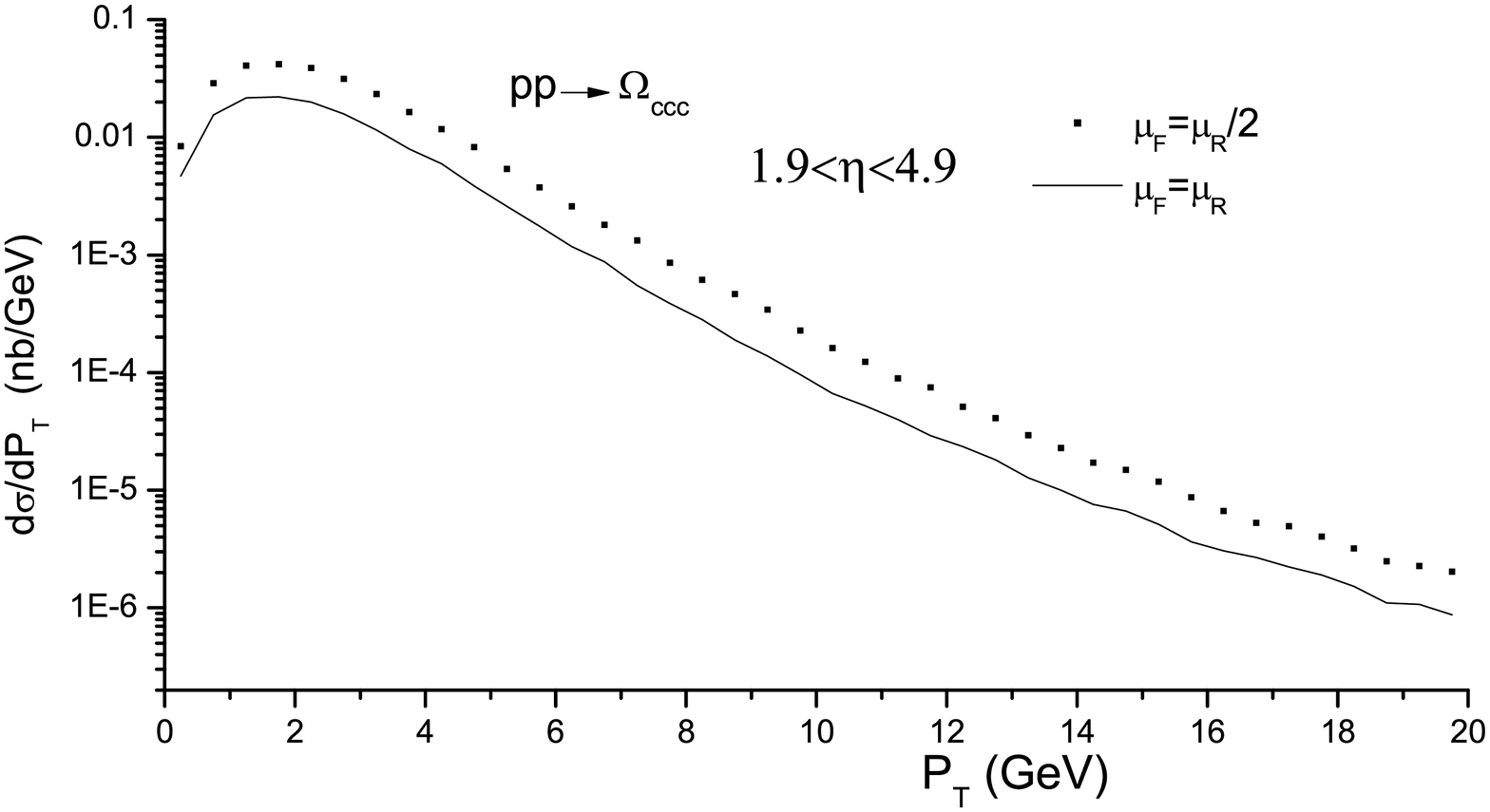}
\includegraphics[width=9cm]{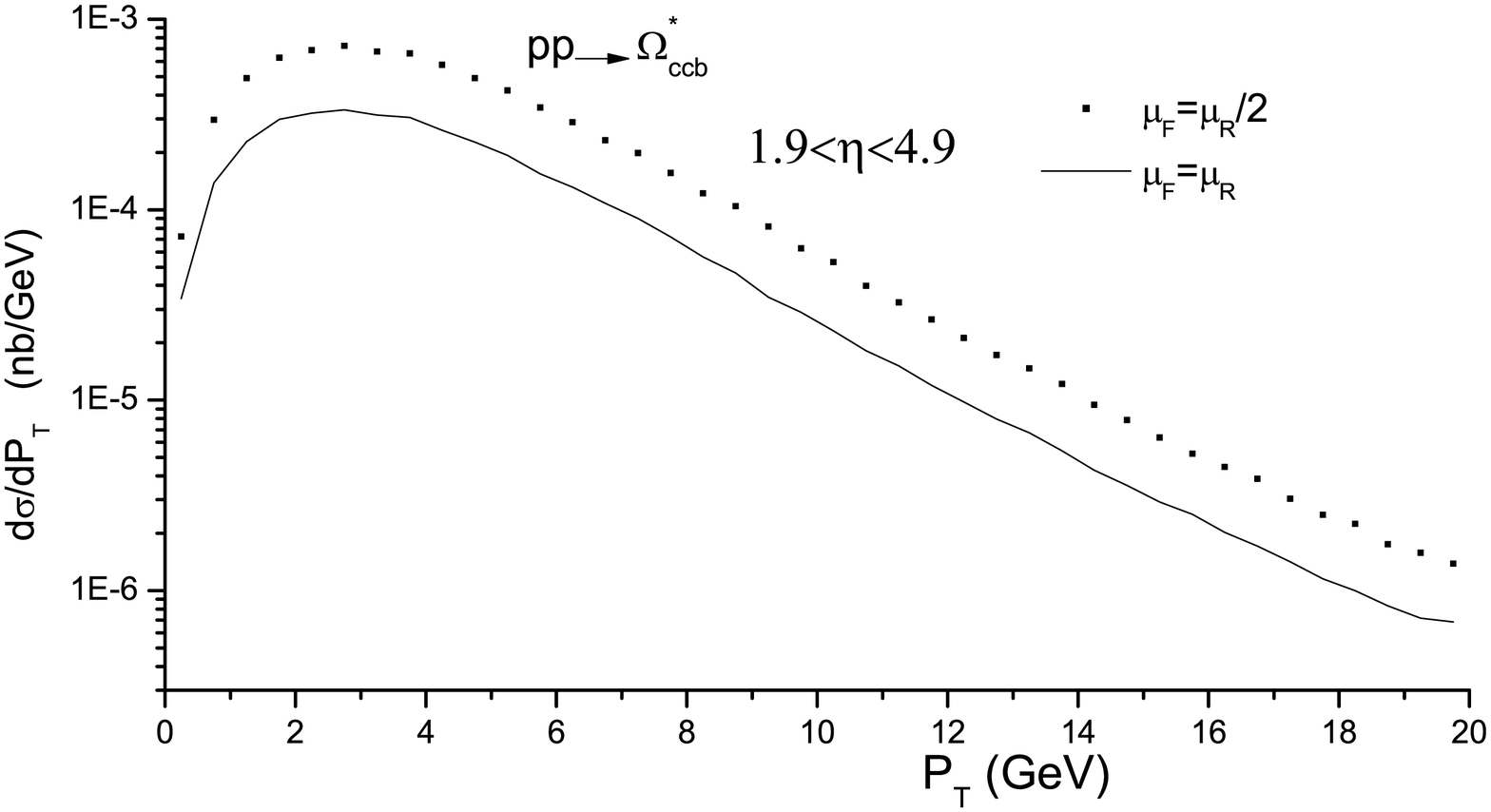}
\includegraphics[width=9cm]{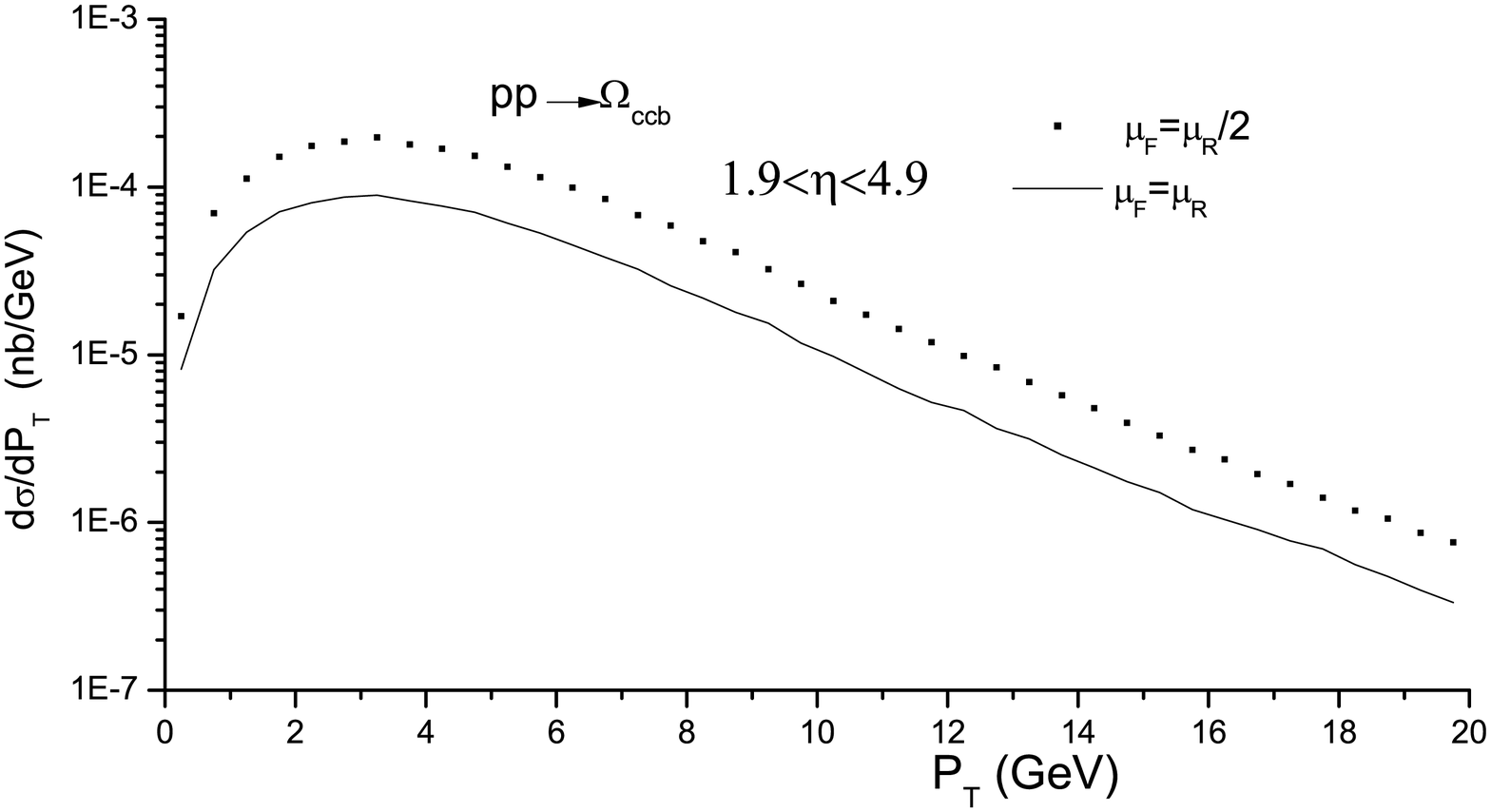}
\caption{The predicted $P_T$-distributions of the triply heavy
baryon production in LHCb, with $\sqrt{S}$=14 TeV. }\label{ptlhc}
\end{figure}

\begin{figure}
\centering
\includegraphics[width=9cm]{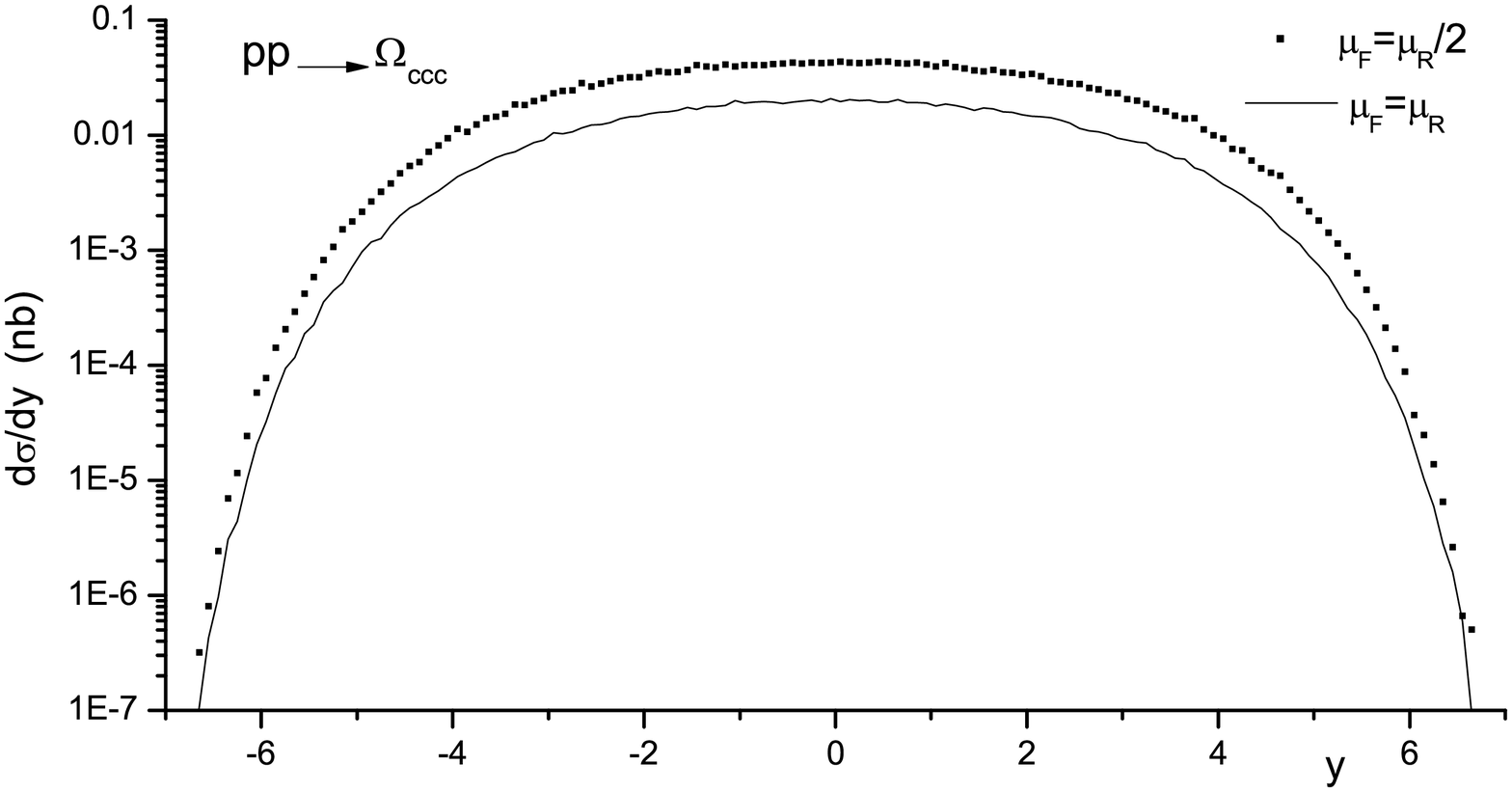}
\includegraphics[width=9cm]{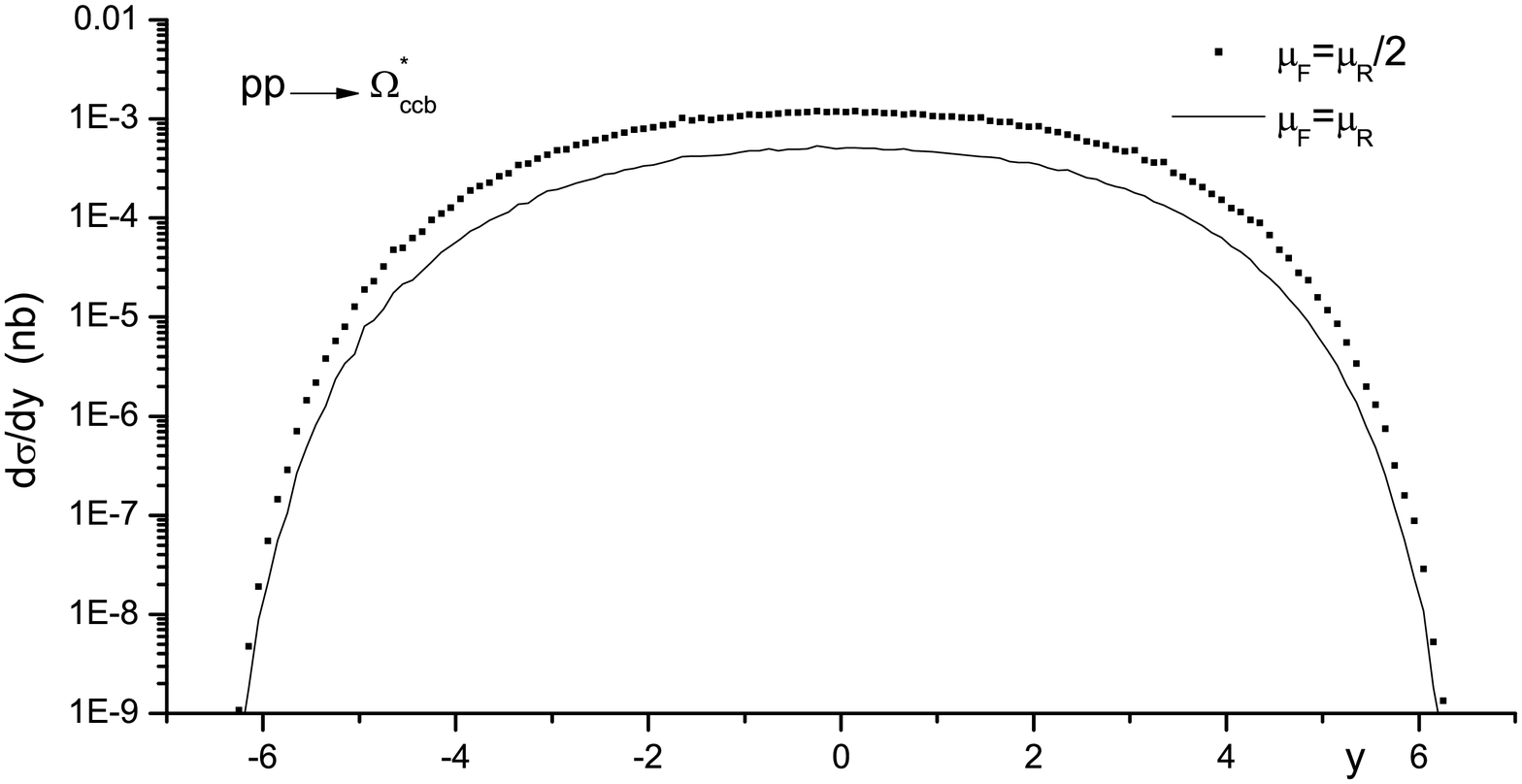}
\includegraphics[width=9cm]{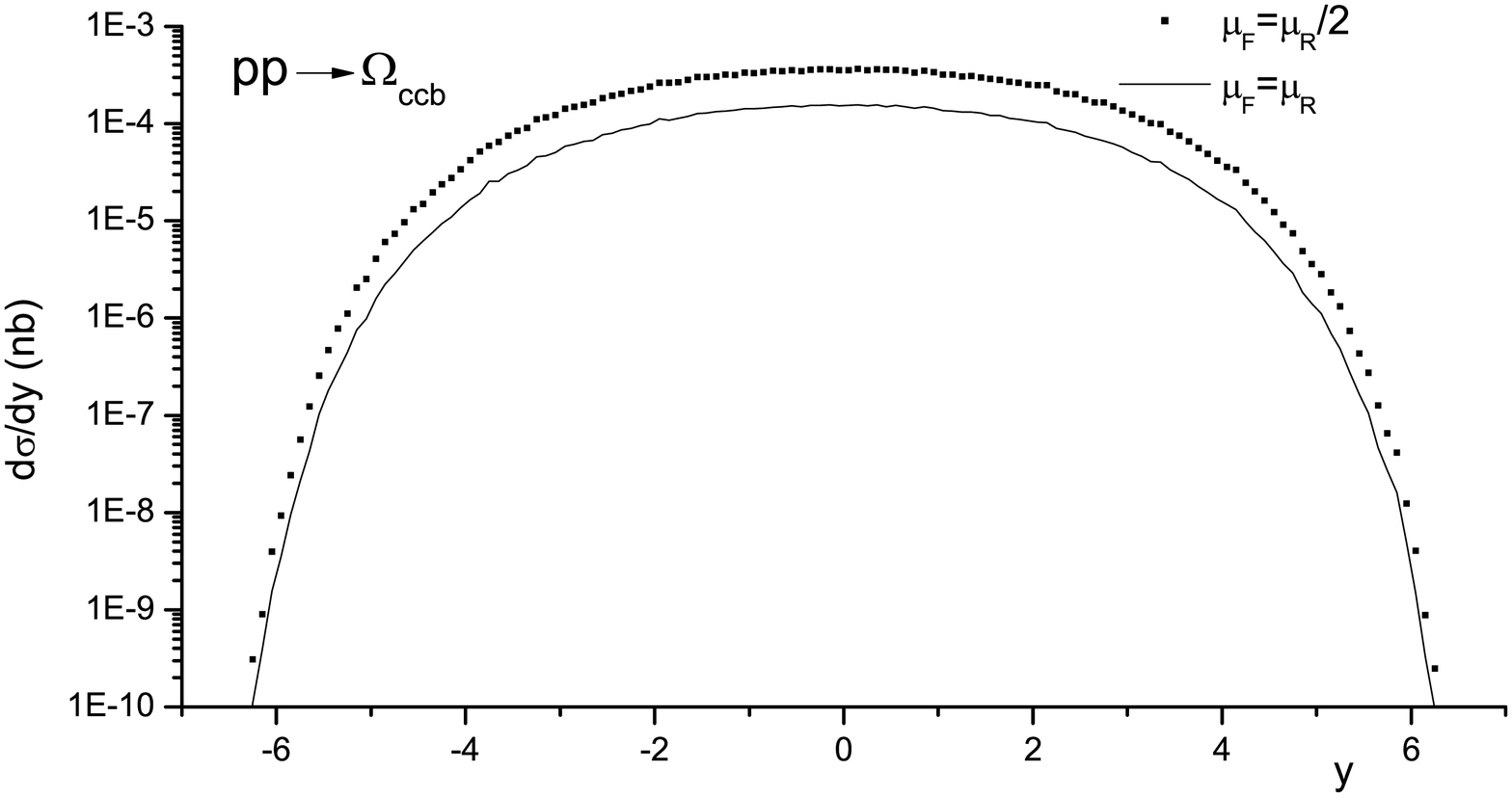}
\caption{The predicted rapidity distributions of the triply heavy
baryon production at LHC with $\sqrt{S}$=7 TeV. }\label{eta:7tev}
\end{figure}

\begin{figure}
\centering
\includegraphics[width=9cm]{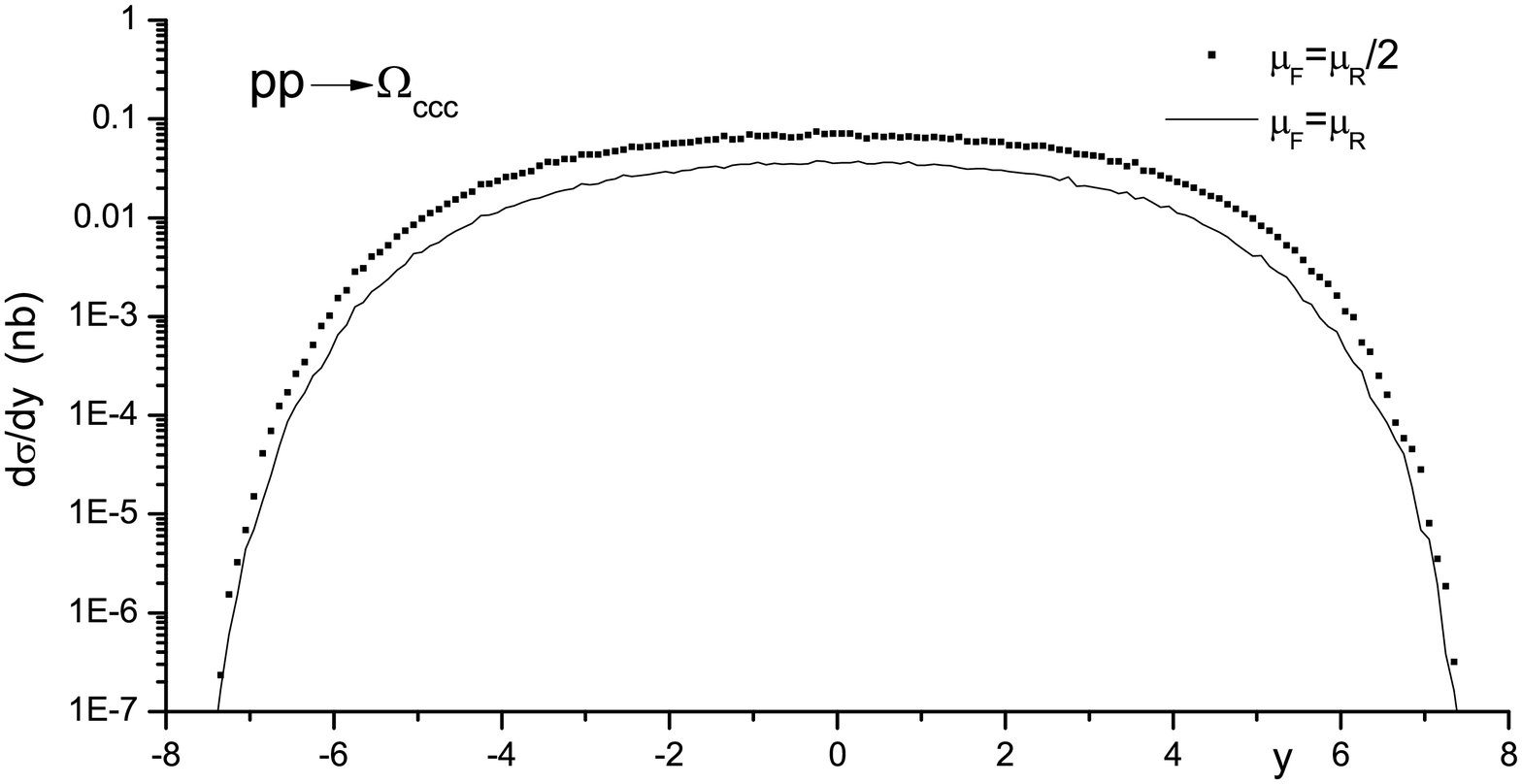}
\includegraphics[width=9cm]{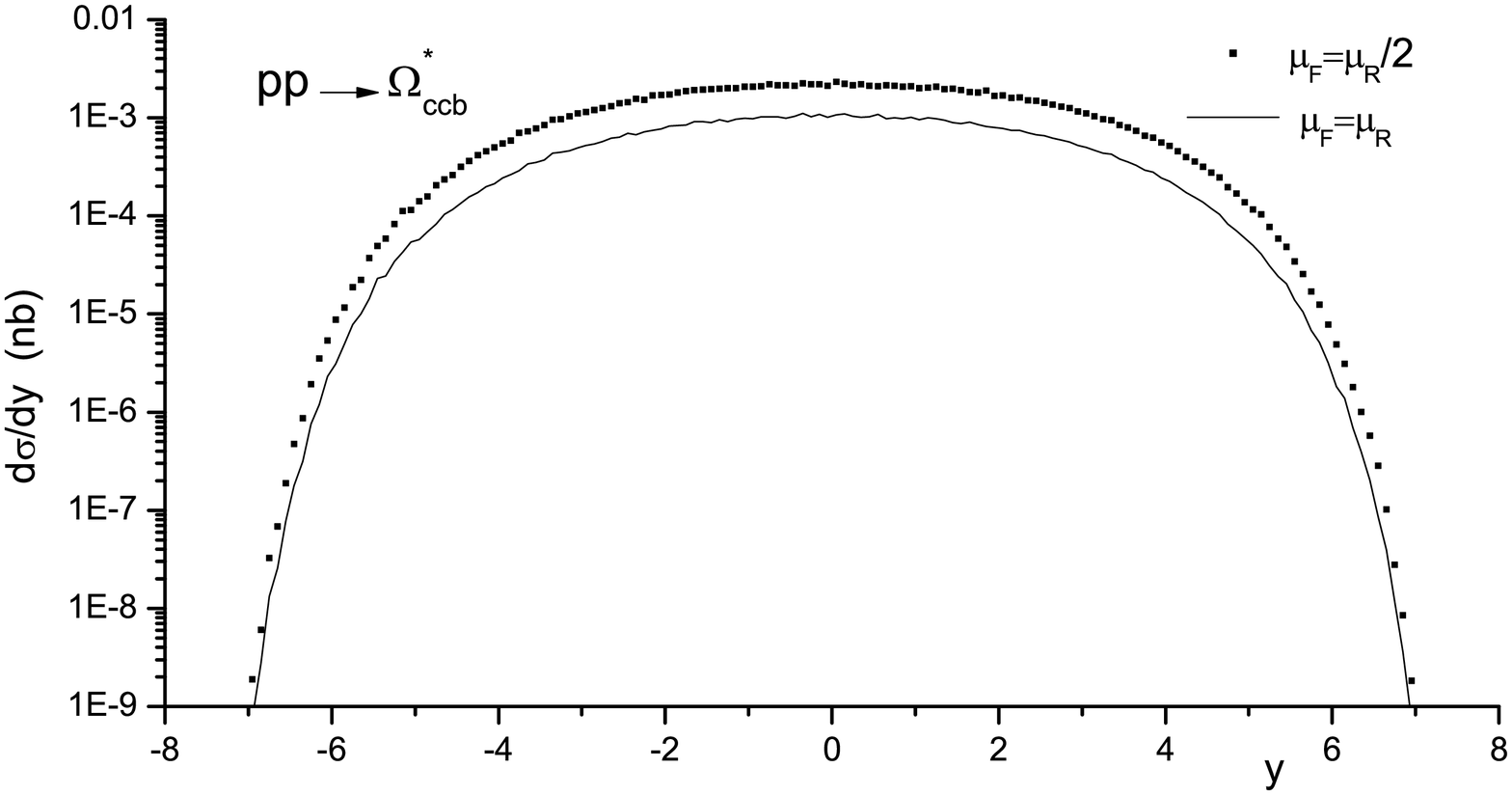}
\includegraphics[width=9cm]{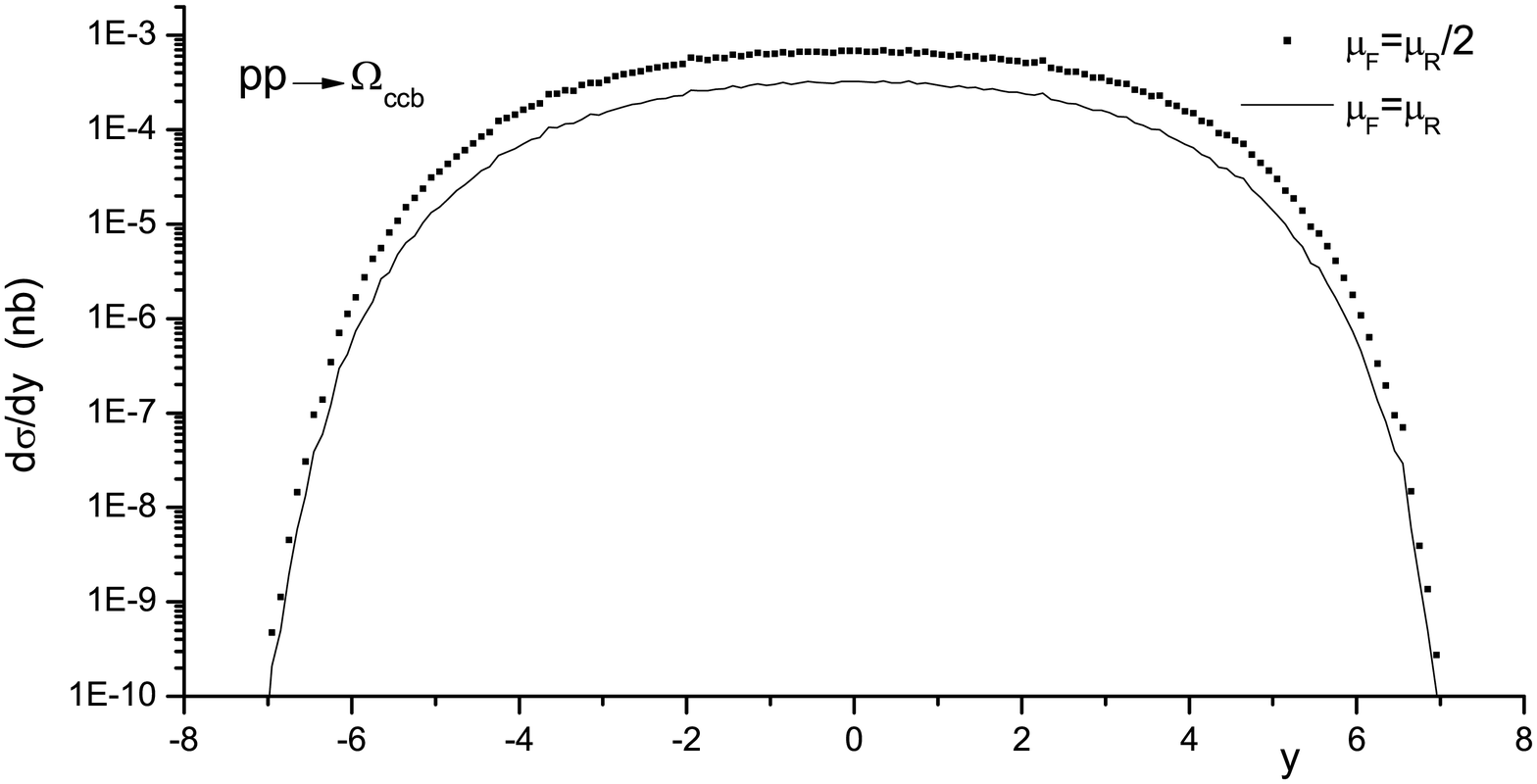}
\caption{The predicted rapidity distributions of the triply heavy
baryon production at LHC with $\sqrt{S}$=14 TeV.}\label{eta}
\end{figure}

We show the $P_T$-distributions and  the rapidity distributions of
the productions of the triply heavy baryons in Figs.
\ref{ptcms:7tev}$-$\ref{eta} at LHC with various LHC running
energies and in various detectors. The differential cross section
drops fast with increasing $P_T$ as expected. Some of the parts in
the $P_T$ distribution are not smooth due to Vegas simulation error
which can be reduced by increasing the event numbers.

It is instructive to look at the fragmentation description for the
production of the triply heavy baryons. One expects that at large
$P_T$ the production is dominated by the fragmentation mechanism.
The production cross sections  are then factored into the product of
the cross sections of the heavy quark pair production and the
fragmentation function. As dimensionless quantities, the QCD leading
order fragmentation function and the fragmentation probability must
be proportional to $\alpha_s^4 \,|\Psi(0,0)|^2 /m^6 $. Since the
process is complicated, one cannot expect to have a simpler analytic
expression.

Comparing our predicted results for the $\Omega_{ccc}$ production
with those fragmentation model descriptions presented in
\cite{frag-1,frag-2}, we see that our numerical result is around 2-3
orders of magnitude higher than that given in \cite{frag-1} and
around 2 orders of magnitude smaller than that given in
\cite{frag-2} for $P_T >10$ GeV. These large discrepancies  cannot
be reconciled by varying parameters such as the heavy quark mass,
factorization energy scale, type of the parton distribution
function, and the rapidity cuts used in the calculations.

To look for the reasons leading to so huge discrepancies between our
result and those in \cite{frag-1,frag-2}, we first look at the
nonperturbative parameter used in the calculations. In
\cite{frag-2}, the authors introduced a nonperturbative
dimension-one parameter $f_B$ as used in the meson production when
they calculated the amplitude of the production of the triply heavy
baryon. From Eq.(5) in Ref. \cite{frag-2}, the nonperturbative
parameter $f_B$ they introduced can be related to our
$\Psi_{ccc}(0,0)$ by $|\Psi_{ccc}(0,0)|^2 \sim f_B^2 \cdot
M_{\Omega_{}}\, m_c^3 $. In their calculation, they took $f_B
^2=0.0625$ GeV$^2$ corresponding to a value of the
$|\Psi_{ccc}(0,0)|^2$ in our paper being $0.458$ GeV$^6$, which is
around two orders of magnitude higher than the value we used. From
quark model calculations, this value is certainly over-estimated.
This may explain  the two orders of magnitude discrepancy between
the results in \cite{frag-2} and ours.

 In \cite{frag-1}, the author introduced two nonperturbative
parameters $\psi_{cc}(0)$ and $ \psi'_{ccc}(0)$ describing the wave
function at origin of the double quarks forming the diquark and the
quark and the diquark  forming the baryon, respectively. His
fragmentation function was then proportional to $|\psi_{cc}(0)\cdot
\psi'_{ccc}(0)|^2 = 0.00345$ GeV$^6$, which was comparable to the
nonperturbative  parameter we used in this paper
$|\Psi_{ccc}(0,0)|^2 = 0.0061$ GeV$^6$. Thus the discrepancy between
the results in \cite{frag-1} and ours cannot be attributed to the
nonperturbative parameters used in the calculations. Notice that the
fragmentation probability given in Eq.(8) in \cite{frag-1} is
proportional to a number $A_c$ he introduced.  In the following
equation after Eq.(8) in \cite{frag-1}, the author showed that $A_c$
was a small number $4.2 \times 10^{-3}$ which arose from a
cancelation between two very large numbers (Surprisingly it is a
10-digit cancelation!). This may give explanation for the
discrepancy between our results and those presented in
\cite{frag-1}. Physically, when they calculated the fragmentation
function using the diquark model , probably only tiny contributions
were accounted for.

 We also calculate the production cross sections for the triply
heavy baryons at Tevatron. Our calculations show that there are
around 2 thousands $\Omega_{ccc}$ events and around  3 hundreds
$\Omega_{ccb}$ events accumulated with 5 GeV $P_T$-cut and 0.6
rapidity cuts for 2 $fb^{-1}$ integrated luminosity at Tevatron.
Considering the detection difficulty, we think these numbers are not
enough to discover the triply heavy baryons at Tevatron.

\section{Signatures and summary}

The particular signatures of the triply heavy baryons in the
detectors can be used to reconstruct the triply heavy baryons.  The
ground states of the triply heavy baryons can decay only through the
weak interaction. Naively, the decay width for $\Omega_{ccc}$ is
triple the width of the free $c$ quark while that for $\Omega_{ccb}$
is double the width of the free $c$ quark plus the decay width of
the free $b$ quark. However, one expects that it receives large
corrections both from the destructive interference between identical
fermions and from large recoil effects. So the total decay widths
may be deviated largely from the naive estimation.

For $\Omega_{ccc}$,  a very interesting non-leptonic  decay mode is
that:
\[
\begin{CD}
\Omega_{ccc}  & &\\
 \downarrow & &\\
 \Omega_{ccs} & \;+\; \pi^+ & \\
\downarrow &  &\\
 \Omega_{css} & \;+\; \pi^+&  \\
\downarrow & &\\
 \Omega_{sss} &\;+\; \pi^+& \bigskip  \\
 \hline \\
\Omega_{ccc} &\to \Omega_{sss} &\;+\; 3\,\pi^+\,.
\end{CD}
\]
Thus with this cascade decay mode, the $\Omega_{ccc}$ finally decays
to $ \Omega_{sss} \;+\; 3\,\pi^+$, which are stable charged
particles in the detectors.  Combining  the tracks of the charged
particles and the invariant mass, the triply heavy baryons can
easily be reconstructed.

For $\Omega_{ccb}$,  the similar but complicated interesting
non-leptonic decay mode is that:

\[
\begin{tabular}{cccccc}
&& \multicolumn{2}{c}{ $\Omega_{ccb} $} &&\\
 && $\swarrow$& \multicolumn{2}{c}{$\searrow $} &\\
 &\multicolumn{2}{c}{$\Omega_{bcs}\;+\;\pi^+$} & &
 $\Omega_{ccc}$ & +\; $\pi^-$  \\
 $\!\!\swarrow$ &\;\;& $\searrow$ & &$\downarrow$  &
  \\
 $\Omega_{bss}$ &$+\; \pi^+$&
 $\Omega_{ccs}$ &$+\; \pi^-$&
 $\Omega_{ccs}$ &$+\; \pi^+$ \\
 $\downarrow$ &  &$\downarrow$ &  &$\downarrow$ &\\
 $\Omega_{css}$ &$+\; \pi^-$&
 $\Omega_{css}$ &$+\; \pi^+$&
 $\Omega_{css}$ &$+\; \pi^+$ \\
$\downarrow$ & &$\downarrow$ &  &$\downarrow $ &\\
 $\Omega_{sss}$ &$+\; \pi^+$&
 $\Omega_{sss}$ &$+\; \pi^+$&
 $\Omega_{sss}$ &$+\; \pi^+$
  \\
 \hline \\
 \multicolumn{4}{c}{ $\Omega_{ccb} \;\to\; \Omega_{sss}\; +\;3\pi^+
\;+\;\pi^- $} &&
\end{tabular}
\]

 Thus with these cascade decay modes, the $\Omega_{ccb}$ final
decay products are $ \Omega_{sss} + 3\,\pi^+ +\pi^- $, which are
stable charged particles in the detectors.  Again combining  the
tracks of the charged particles and the invariant mass, the triply
heavy baryons can easily be reconstructed.

We can also analyze the semileptonic decays of  the $\Omega_{ccc}$,
$\Omega_{ccb}$. The cascade decays are  $c\to s$ or $b \to c \to s
$, with emitting leptons and neutrinos. This cascade decay feature
with the decay vertex  can also be used to identify the triply heavy
baryons. Moreover, one can also combine the semileptonic decay with
the nonleptonic decay to reconstruct the triply heavy baryons.

In conclusion, our results show that a number of triply heavy
baryons events can be accumulated at LHC. They can be reconstructed
with their unique signatures in detectors.
 Triply heavy baryons are
very interesting hadrons to be explored for they provide particular
information about strong interaction, hadron structure, and weak
decay of heavy baryons.
They are still absent in the particle data booklet after the heavy
quarkonium has been discovered over three decades. Our results show
that it is quite promising to discover those triply heavy baryons in
LHC experiments both for large number events and for their unique
signatures in detectors. One may be waiting for  an exciting time to
discover them at LHC.

\newpage
\section{Acknowledgements}
One of the authors (Su-Zhi Wu) wants to thank Bin Gong for great
help for matching the color factors and helpful discussions. This
work is partly supported  by the NSFC with contract number 10875156.

\newpage \section{Appendix}
\subsection{Nonrelativistic states}\label{non-baryon}
The $S-$ wave state of the nonrelativistic triply heavy baryon in
its rest frame is constructed as:
\begin{eqnarray} \label{3cnrqcd}
|\Omega_{Q_1Q_2Q_3},s,s_Z\rangle=\sqrt{2\overline{M}}\int\frac{d^3\overrightarrow{V}_1}{(2\pi)^3}\frac{d^3\overrightarrow{V}_2}{(2\pi)^3}
\sum_{\xi_1,\xi_2,\xi_3}\sum_{\eta_1,\eta_2,\eta_3}\frac{\varepsilon^{\xi_1\xi_2\xi_3}}{\sqrt{6}}
\langle
s,s_Z|\eta_1,\eta_2,\eta_3\rangle\nonumber\\
\frac{1}{\sqrt{2E_12E_22E_3}}\frac{1}{\sqrt{d!}}\psi(\overrightarrow{V}_1,\overrightarrow{V}_2)
\times|Q_1,\xi_1,\eta_1,\vec{p}_1=\overrightarrow{V}_1\rangle\nonumber\\
\times|Q_2,\xi_2,\eta_2,\vec{p}_2=\overrightarrow{V}_2\rangle
\times|Q_3,\xi_3,\eta_3,\vec{p}_3=-\overrightarrow{V}_1-\overrightarrow{V}_2\rangle\; ,\nonumber\\
\end{eqnarray}
where, $\xi_i$, $\eta_i$, $m_i$, and ($E_i$, $\vec{p}_i$)
($i=1,2,3$) are the color, the spin, the mass, and the four-momentum
of the heavy quark $Q_i$; $V_1$ and $V_2$ are the relative momenta
among the heavy quarks; $\langle s,s_Z|\eta_1,\eta_2,\eta_3\rangle$
is the C-G coefficient; $s$ and $s_Z$ are the spin of the baryon and
its third component, respectively;
$\psi(\overrightarrow{V}_1,\overrightarrow{V}_2)$ is the wave
function of the baryon in the momentum space.

The quark states are normalized by:
\begin{eqnarray} \label{nor-q}
\langle f^{'},\xi^{'},\eta^{'},\vec{p}^{'}|f,\xi,\eta,\vec{p}
\rangle=\delta_{f^{'}f}\delta_{\xi^{'}\xi}\delta_{\eta^{'}\eta}(2\pi)^32E_f\delta^{(3)}(\vec{p}^{'}-\vec{p})\;,
\end{eqnarray}
where $f^{'}$ and $f$ are the heavy flavors.

$\psi(\overrightarrow{V}_1,\overrightarrow{V}_2)$ is normalized by:
\begin{eqnarray} \label{nor-wa}
\int\frac{d^3\overrightarrow{V}_1}{(2\pi)^3}\frac{d^3\overrightarrow{V}_2}{(2\pi)^3}\psi^*(\overrightarrow{V}_1,\overrightarrow{V}_2)\psi(\overrightarrow{V}_1,\overrightarrow{V}_2)=1\;.
\end{eqnarray}
By the Eqs. (\ref{nor-q}) and (\ref{nor-wa}), we know the baryon is
normalized as:
\begin{eqnarray} \label{nor}
\langle \Omega,s^{'},s^{'}_Z,\vec{P}^{'}|\Omega,s,s_Z,\vec{P}
\rangle=(2\pi)^32E\delta^{s,s^{'}}\delta^{s_Z,s^{'}_Z}\delta^{(3)}(\vec{P}^{'}-\vec{P})\;
.
\end{eqnarray}
The wave function at the origin  is related to
$\psi(\overrightarrow{V}_1,\overrightarrow{V}_2)$ by:
\begin{eqnarray} \label{wave}
\Psi(0,0)=\int\frac{d^3\overrightarrow{V}_1}{(2\pi)^3}\frac{d^3\overrightarrow{V}_2}{(2\pi)^3}\psi(\overrightarrow{V}_1,\overrightarrow{V}_2)\;
.
\end{eqnarray}

\subsection{Spinor and polarization vector }\label{sec-3}

For on-shell quark or antiquark with 4-momentum
$p^{\mu}=(p^0,p^1,p^2,p^3)$ and mass $m$ satisfying $p^2=m^2$, the
spinor of the quark reads:
\begin{eqnarray}\label{spin-quark}
u^{\frac{1}{2}}(p)&=&(f_1(|\vec{p}|+p^3,p^1+i p^2),\;
f_2(|\vec{p}|+p^3,p^1+i
p^2))\; , \nonumber \\
u^{-\frac{1}{2}}(p)&=&(f_2(-p^1+i p^2,|\vec{p}|+p^3),\; f_1(-p^1+i
p^2,|\vec{p}|+p^3))\; ,
\end{eqnarray}
and the spinor of the antiquark reads:
\begin{eqnarray} \label{spin-antiquark}
v^{\frac{1}{2}}(p)&=&(-f_2(-p^1+i p^2,|\vec{p}|+p^3)\;,\; f_1(-p^1+i
p^2,|\vec{p}|+p^3))\;, \nonumber \\
v^{-\frac{1}{2}}(p)&=&(f_1(|\vec{p}|+p^3,p^1+i p^2)\;,\;
-f_2(|\vec{p}|+p^3,p^1+i p^2))\;,
\end{eqnarray}
where
\begin{eqnarray}
f_1=\frac{\sqrt{p^0-|\vec{p}|}}{\sqrt{2|\vec{p}|(|\vec{p}|+p^3)}}\;,
&\;\;&
f_2=\frac{\sqrt{p^0+|\vec{p}|}}{\sqrt{2|\vec{p}|(|\vec{p}|+p^3)}}\;.
\end{eqnarray}

Setting the beam line as the Z-axis,  two physical polarization
vectors of the gluon are:
$$\epsilon^+_{\mu}=(0,1,0,0)\; , \ \ \ \ \ \ \ \ \epsilon^-_{\mu}=(0,0,1,0)\; .$$
\subsection{The $\lambda$ matrices and the Feynman
rules}\label{sec-2} In our calculations, the $\lambda$
matrices of $SU(N_c)$  satisfy:\\
$$ [\lambda^a,\lambda^b]=if^{abc}\lambda^c \;, \ \
tr[\lambda^a,\lambda^b]=\delta^{ab}\;,$$
$$
\sum_{a}(\lambda^a)_{ij}(\lambda^a)_{kl}=\delta_{il}\delta_{jk}-\frac{1}{N_c}\delta_{ij}\delta_{kl}\;
,$$
where $N_c=3$ in QCD.\\

Feynman rules for those vertex are given as follows.\\
Quark-antiquark-gluon vertex:\\
$$i\frac{g}{\sqrt{2}}\gamma_{\mu}(\lambda^a)_{ij}\; ,$$
Tri-Vector vertex in order (123) with all momenta incoming to vertex:\\
$$\frac{g}{\sqrt{2}}f^{a_{1}a_{2}a_{3}}[(p_{1}-p_{2})_{\mu_{3}}g_{\mu_{1}\mu_{2}}+(p_{2}-p_{3})_{\mu_{1}}g_{\mu_{2}\mu_{3}}+(p_{3}-p_{1})_{\mu_{2}}g_{\mu_{1}\mu_{3}}]\; ,$$
Quadruple-Vector vertex:
\begin{center}
      \includegraphics[width=0.32\textwidth]{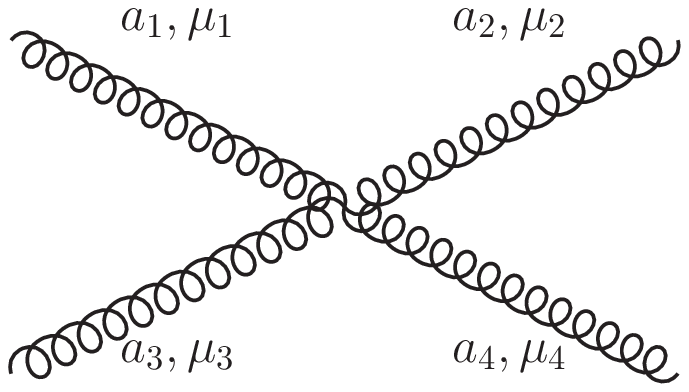}
\end{center}
$$-i\frac{g^2}{2}[f^{a_{1}a_{2}e}f^{a_{3}a_{4}e}(g^{\mu_{1}\mu{3}}g^{\mu_{2}\mu{4}}-g^{\mu_{1}\mu{4}}g^{\mu_{2}\mu{3}})$$
$$+f^{a_{1}a_{3}e}f^{a_{2}a_{4}e}(g^{\mu_{1}\mu{2}}g^{\mu_{3}\mu{4}}-g^{\mu_{1}\mu{4}}g^{\mu_{2}\mu{3}})+f^{a_{1}a_{4}e}f^{a_{2}a_{3}e}(g^{\mu_{1}\mu{2}}g^{\mu_{3}\mu{4}}-g^{\mu_{1}\mu{3}}g^{\mu_{2}\mu{4}})]\; .$$

\end{document}